\documentclass[lettersize,journal]{IEEEtran}
\IEEEoverridecommandlockouts
\usepackage{cite}
\usepackage{caption}

\usepackage{placeins}

\usepackage{cuted}

\usepackage{float}
\usepackage{amsmath,amssymb,amsfonts}
\usepackage{algorithm}
\usepackage{algorithmicx}
\usepackage{algpseudocode}
\usepackage{graphicx}
\usepackage{textcomp}
\usepackage{xcolor}
\colorlet{blue}{black}
\usepackage[left=0.76in, right=0.65in, top=0.76in, bottom=1.05in]{geometry}

\DeclareSymbolFont{Xlargesymbols}{OMX}{cmex}{m}{n}
\DeclareMathSymbol{\Xsum}{\mathop}{Xlargesymbols}{80}

\title{Waveguide to Meaning: Semantic-Aware NOMA for Pinching-Antenna Systems}
\author{\IEEEauthorblockN{Ishtiaque Ahmed, Haris Pervaiz, and Leila Musavian}
\thanks{The authors are with the School of Computer Science and Electronic Engineering, University of Essex, Wivenhoe Park, Colchester CO4 3SQ, United Kingdom (e-mail: { \{ishtiaque.ahmed, haris.pervaiz, leila.musavian\}@essex.ac.uk). This work has been submitted to the IEEE for possible publication. Copyright may be transferred without notice, after which this version may no longer be accessible.}

This work was supported by the UK Research and Innovation under the UK government’s Horizon Europe funding guarantee through MSCA-DN SCION Project Grant Agreement No.101072375 [grant number: EP/X027201/1].} 
 }
    
\begin{document}
\maketitle

\begin{abstract}
We investigate the performance of the pinching-antenna systems (PASS) for semantic communication (SC) in both single-waveguide and multi-waveguide scenarios, under the constraints of bit-user quality of service (QoS) and bit-to-semantic decoding order in a heterogeneous users downlink non-orthogonal multiple access (NOMA). Multiple pinching antennas in the single-waveguide scenario are at a minimum adjacent spacing required to prevent mutual coupling. An alternating optimization (AO)-based algorithm optimizes users power allocation coefficients and position of pinching antennas in the single-waveguide NOMA framework. For the multi-waveguide scenario, assuming adjacent waveguides at a sufficient lateral distance apart, the waveguides power allocation subproblem is solved using monotonic optimization and minorization-maximization (MM) approach. Specifically, a lower bound surrogate is iteratively maximized under the feasibility constraints such that a non-decreasing sequence of objective is obtained. Numerical results demonstrate that the NOMA based PASS exploiting SC offer higher semantic spectral efficiency (SE) while fulfilling the bit-user QoS when compared to the considered conventional fixed antenna system. Notably, the multi-waveguide scenario becomes more beneficial for creating adjustable wireless channels in stringent conditions with higher bit-user QoS and wider coverage area requirements.
\end{abstract}
\vspace{-1em}

\vspace{-1em}
\section{Introduction} \label{intro}
Next-generation networks are expected to support intelligent communication for diverse tasks over shared time-frequency resources \cite{wang2024adaptive}. Classical multi-antenna systems have resulted in performance improvements compared to single antenna systems, but they typically exploit fixed antennas for bit-based communication \cite{zeng2016millimeter}.

With recent advances in multiple-input multiple-output (MIMO) techniques for enabling high-speed and massive machine-to-machine sixth-generation (6G) communications, flexible-antenna systems \cite{ma2023mimo} and intelligent semantic communication (SC) \cite{letaief2019roadmap} are expected to enhance system capacity. Flexible-antenna architectures have recently emerged as a compelling solution, turning the propagation channel into a controllable resource without any costly radio frequency (RF) chains. Among them, the pinching-antenna systems (PASS) stand out in creating strong line-of-sight (LoS) links due to its ability to flexibly place dielectric mediums over the waveguide \cite{ding2025flexible}. \color{blue} Unlike conventional fixed-location antenna systems, PASS provide continuous spatial reconfigurability by allowing the radiating points to be adjusted along the dielectric waveguide. These pinching locations affect both the free-space spherical-wave propagation distance and the guided-wave phase accumulated inside the waveguide. Therefore, PASS enable pinching beamforming, where the radiating locations are optimized according to the users’ spatial geometry. \color{black} Moreover, PASS enable new MIMO and non-orthogonal multiple access (NOMA) integration with one RF chain to feed multiple spatially distributed apertures, without additional hardware overhead. These traits make PASS a compelling 6G building block for next generational multiple access (NGMA) and dense urban deployments \cite{yang2025pinching}. Foundational studies on PASS predominantly focused on single-waveguide where multiple pinching antennas share the same RF and their adjacent distances respect a minimum-spacing rule to avoid inter-element coupling. Lately, pinching antennas have been introduced to multi-waveguide architectures \cite{liu2025pinching, hu2025sum, fu2025power}. \color{blue} Notably, multi-waveguide PASS offer additional spatial and power-domain degrees of freedom. Each waveguide contributes an independently guided-radiated path to the received signal, and the effective channel is determined by coherent combining across waveguides. With the optimal pinching locations and waveguide power-splitting, the multi-waveguide architecture exploits near-field spatial focusing and coherent combining beyond the single-waveguide case.

\color{black} Recent PASS studies have profiled their gains with design principles that quantify attenuation and highlight spacing to mitigate coupling \cite{xiao2025channel} and joint transmit–pinching beamforming \cite{xu2026joint} with a uniform RF over the waveguide. \color{blue} Attenuation-aware PASS require an additional guided-path loss factor, which is particularly important for long waveguides, high-loss guiding structures, or large service regions \cite{1-1a, wang2025antenna, 1-1c}. \color{black} The work in \cite{tyrovolas2025performance} provides a detailed analysis on outage probability and average rate in PASS by taking into account the waveguide losses. Sum-rate maximization for traditional bit communication has been done in \cite{zhou2025sum}, providing closed-form power allocation and pinching antennas placement on a single-waveguide. It also highlights the trade-off between phase accuracy and path loss due to antenna repositioning. Beyond single-user bit-rate maximization, PASS have been combined with NOMA via power-domain multiplexing for increasing spectral efficiency (SE) and reducing outage probability \cite{wang2025antenna, wang2025antenna2}. Analytical and algorithmic frameworks have been developed for antenna activation, sum-rate maximization, and power minimization subject to quality of service (QoS) and spacing constraints \cite{ding2025analytical}. Authors in \cite{shan2025exploiting} investigated PASS for multicast communications and devised a majorization-minimization approach along with the alternating optimization (AO) to optimize the transmit and pinching beamformers iteratively in a multi-waveguide scenario. Moreover in \cite{zhou2025gradient}, a gradient-based data-driven optimization offers substantial sum-rate improvement over conventional AO in large multi-waveguide configurations. These works demonstrate that appropriate spacing and pinching position in PASS architectures enlarge the near–far channel gain effects for enabling successive interference cancellation (SIC) in the conventional communication paradigm. 

Traditional bit communication is bounded by Shannon capacity and relies on mutual information in the entropy domain, which is linked to the technical level \cite{shannon1998mathematical}, while overlooking the semantic and effectiveness levels. With the proliferation of wireless subscribers and the requirement of intelligent devices to meet the next-generational communication requirements, investigation of the second and third levels of communication is inevitable \cite{letaief2019roadmap}, \cite{gunduz2022beyond}. On the other hand, SC targets efficiency with transmission of contextual meaning rather than raw bits \cite{yang2022semantic}. Studies suggest that SC offers superior performance under low signal-to-noise ratio (SNR) conditions, where mutual information based communication becomes susceptible to noise \cite{han2022semantic}, \cite{ahmed2025semantic}. At higher SNR regimes, SC yields diminishing performance suggesting that it should be complemented with conventional communication for an enhanced overall performance \cite{yan2022resource}. However, the design principle of bit-to-semantic decoding \cite{mu2022heterogeneous} must be adopted due to the pre-trained neural network architecture in SC \cite{chen2023uplink}.

With the aid of certain neural network architectures, state-of-the-art SC works by extracting only the most important features in any message for its transmission and reconstruction to enhance SE. Expressed in \cite{yan2022resource}, the semantic information rate is as
\begin{equation}
R_{\text{S}} = \frac{WI}{KL} \epsilon_K(\gamma),
\label{semeq}
\end{equation}
where $\epsilon_K(\gamma)$ is the sigmoid-shaped semantic similarity function whose value ranges between zero and one [16], $\gamma$ represents the received SNR, $W$ is the channel bandwidth, $I$ is the amount of semantic information in any message with semantic units (suts), $K$ represents the average number of semantic symbols transmitted for each word, and $L$ denotes the number of words in a sentence. For a unit bandwidth, the resulting $R_{\text{S}}$ is therefore measured in suts/s/Hz.

A joint source-channel coding framework, namely DeepSC provides a practical text SC transceiver \cite{xie2021deep}. DeepSC preserves semantic fidelity under low-SNR for the fading channels and is equipped with neural networks \cite{getu2025semantic}. Its transformer-based encoder \cite{devlin2018bert} maps each sentence to a compact sequence of $K$ semantic symbols, which are then directed for transmission. At the receiver, a decoder reconstructs the sentence by maximizing $\epsilon_K(\gamma)$, increasing the similarity between the output and source texts. Moreover, the DeepSC sigmoid shaped curve suggests diminishing returns at high SNR but large gains in low-to-moderate SNR regime.

\color{blue} It is important to distinguish the adopted semantic rate from the Shannon rate. The Shannon rate measures the maximum reliable bit information rate over a channel. In contrast, the semantic rate given above measures the amount of meaningful semantic information reconstructed by a trained DeepSC. If the semantic objective were replaced by the Shannon rate \(\log_2(1+\gamma)\), the qualitative design direction would remain similar in the considered single-semantic-user setting, because both objectives are monotonically increasing functions of \(\gamma\). However, the resulting objective would quantify bit-level SE rather than providing interpretation in terms of meaningful semantic information reconstruction.

\color{blue} Specifically, the adopted semantic similarity approximation at a given SNR is introduced to enable tractable analysis of the pre-trained semantic transceiver \cite{xie2021deep} for a given text-transmission task, dataset, neural architecture and number of semantic symbols per word $K$. In text-based SC, the transceiver extracts semantic features from the input sentence and maps them into semantic symbols, while the receiver reconstructs the sentence through the channel decoder and semantic decoder. The original and recovered sentences are then compared using a semantic similarity metric based on bidirectional encoder representations from transformers (BERT) \cite{devlin2018bert}. Let $s$ and $\hat{s}$ denote the original and recovered sentences, respectively. The semantic fidelity is measured through a sentence-level similarity metric as \cite{xie2021deep}
\begin{equation}
\textnormal{match}(s, \hat{s})=\frac{B(s) B(\hat{s})^T}{\|B(s)\|\|B(\hat{s})\|},
\end{equation}
where $B(\cdot)$ denotes the sentence embedding generated by BERT. The metric $\textnormal{match}(s, \hat{s})$ takes values in the range [0,1], where a larger value indicates higher semantic similarity between the transmitted and recovered sentences. For a trained DeepSC, the average semantic similarity depends on $\gamma$ and $K$ \cite{mu2023exploiting}. Therefore, we have
\begin{equation}
\textnormal{match} \approx \epsilon_K(\gamma) .
\end{equation}
According to \cite{yan2022resource}, the value of $\epsilon_K(\gamma)$ under different $K$ and $\gamma$ can be obtained by running the DeepSC tool. However, the theoretical investigation on the output of DeepSC is non-trivial due to lack of an explicit form. Existing SC works approximate $\epsilon_K(\gamma)$ using a generalized logistic function because the measured semantic similarity is bounded, monotonically non-decreasing with SNR, and exhibits an S-shaped saturation behaviour \cite{mu2022heterogeneous}, \cite{xie2021deep}, \cite{mu2023exploiting}. Therefore, we clarify the adopted model as
\begin{equation}
\epsilon_K\left(\gamma , \boldsymbol{\theta}_K\right)=A_{K, 1}+\frac{A_{K, 2}-A_{K, 1}}{1+e^{-\left(C_{K, 1} \gamma+C_{K, 2}\right)}},
\end{equation}
where $\boldsymbol{\theta}_K=\left\{A_{K, 1}, A_{K, 2}, C_{K, 1}, C_{K, 2}\right\}$. The lower (left) asymptote, upper (right) asymptote, growth rate, and the mid-point parameters of the logistic function are respectively denoted by $A_{K,1}$, $A_{K,2}$, $C_{K,1}$, and $C_{K,2}$. These parameters are obtained by fitting the DeepSC semantic-similarity samples using a data-regression method \cite{mu2022heterogeneous}, typically under a minimum mean-square-error criterion to make the semantic-similarity expression analytically tractable.

The semantic rate can therefore be written as
\begin{equation}
R_\text{S}=\frac{W I}{K L} \epsilon_K\left(\gamma , \boldsymbol{\theta}_K\right).
\end{equation}
If a minimum semantic fidelity is required, the effective semantic rate can be written as
\begin{equation}
R_\text{S}^{\mathrm{eff}}=\frac{W I}{K L} \epsilon_K\left(\gamma , \boldsymbol{\theta}_K\right) \mathbf{1}\left[\epsilon_K\left(\gamma , \boldsymbol{\theta}_K\right) \geq \epsilon_{\mathrm{th}}\right] .
\end{equation}
Here, $\mathbf{1}[\cdot]$ is a binary indicator function,whose value is 1 if $\left[\epsilon_K\left(\gamma , \boldsymbol{\theta}_K\right) \geq \epsilon_{\mathrm{th}}\right]$; and the value is 0, otherwise.

\color{black} 
Works in \cite{mu2023exploiting, meng2023multi, ahmed2025hybrid} adopt the DeepSC approach for enabling heterogeneous users coexistence under NOMA. These studies numerically show that semantic transmission is robust at low-to-moderate SNR and can be naturally paired with SIC-based access. However, to the best of our knowledge, none of the existing studies have yet incorporated flexible PASS to create the channel disparity that NOMA can exploit. \color{blue} The downlink PASS rate-maximization work in \cite{1-12} considers a single-user setting and discusses a two-stage method for pinching locations in which the pinching locations are first optimized to reduce large-scale path loss and then refined to improve constructive phase alignment. However, it does not consider the heterogeneous users transmission where simply moving the pinching antenna toward the semantic user may violate the bit-user rate or SIC constraints. Furthermore, the in-waveguide attenuation-aware PASS work in \cite{1-13} shows that optimizing the pinching antenna position becomes highly involved, as the placement must balance free-space path-loss reduction, in-waveguide attenuation, and phase effects. \color{black}With the dynamic control of antennas placement in PASS, semantic and bit users can be efficiently allocated powers to keep semantic users in the high-slope regime while ensuring the QoS for bit users. Attenuation challenges in PASS can also be overcome with favourable geometric alignment and reduced overhead via SC. On the other hand, NOMA’s potential for 6G lies in efficient superposition coding and SIC to boost SE and connectivity under diverse QoS, as its performance is sensitive to channel disparities and decoding order.

\vspace{-2.0285em}
\subsection{Motivations and Contributions}
Although the application of PASS has gained popularity over the fixed positional conventional antenna system (CAS), their potential for SC has not been addressed. Our work presents a promising research avenue by jointly exploiting PASS mobility and SC within NGMA directions. This work provides an analysis on both the single- and multi-waveguide PASS for a consistent signal model, power constrain, and QoS requirements within a unified heterogeneous users framework. \color{blue} It is worth noting that the adopted communication-theoretic model neglects residual electromagnetic coupling among adjacent pinching antennas and among different waveguides. A fully coupling-aware PASS design would require calibrated coupling coefficients obtained from measurements or full-wave electromagnetic simulations, which is left for future investigation. \color{black} The main contributions of this work are summarized as follows:
\begin{itemize}
\item This paper proposes a PASS-enabled downlink heterogeneous users NOMA framework to maximize the semantic SE under the minimum rate requirement of the bit user. For this, it formalizes both single- and multi-waveguide scenarios in which pinching antennas are deployed to deliver superimposed signals to heterogeneous users within the coverage area.

\item This paper formulates an optimization problem for both scenario types. In the single-waveguide scenario, multiple pinching antennas are deployed on one waveguide, whereas in the multi-waveguide setup, each waveguide carries only one pinching antenna. To avoid coupling effects, adjacent pinching antennas on the same waveguide are at a minimum distance apart, and the inter-waveguide spacing is also selected to avoid coupling between the nearest pinching antennas. This allows for a direct performance comparison in terms of semantic SE for the single- and multi-waveguide architectures. 

\item For the single-waveguide scenario with uniform power across the pinching antennas, we propose an AO-based algorithm that gives optimal power allocation coefficients for users and pinching antenna positions with phase-alignment sensitivity.

\item For the multi-waveguide scenario, the overall semantic SE maximization problem is decomposed into three subproblems. We solve the users power allocation and pinching antenna position subproblems similarly to the single-waveguide setup, except that the fine-scale phase alignment step is only done in the single-waveguide case. We then address the waveguide power allocation subproblem using a minorization-maximization (MM)-based approach, where a surrogate objective function is iteratively maximized to obtain the solution.

\item We adopt and justify the bit-to-semantic decoding order in a heterogeneous users network to guarantee the bit-user QoS requirement while leveraging the channel disparity induced by PASS.

\item We provide numerical results for the single-waveguide and multi-waveguide PASS to validate their effectiveness in a heterogeneous users network. The results show that: i) PASS outperform the CAS in terms of semantic SE, especially with a larger number of pinching antennas under low transmit power and small coverage area; ii) finer phase alignment accuracies in the single-waveguide PASS yield better performance, while tighter phase alignment for the semantic user provides the most advantage for semantic SE improvement; and iii) multi-waveguide PASS becomes more beneficial under stringent conditions, such as higher bit-user rate and wider coverage area requirements.
\end{itemize}

\vspace{-0.9975em}
\section{Single-Waveguide System Model}
\begin{figure}[t]
\centering
\includegraphics[trim={0cm 0cm 0cm 0cm},clip,width=1\columnwidth]{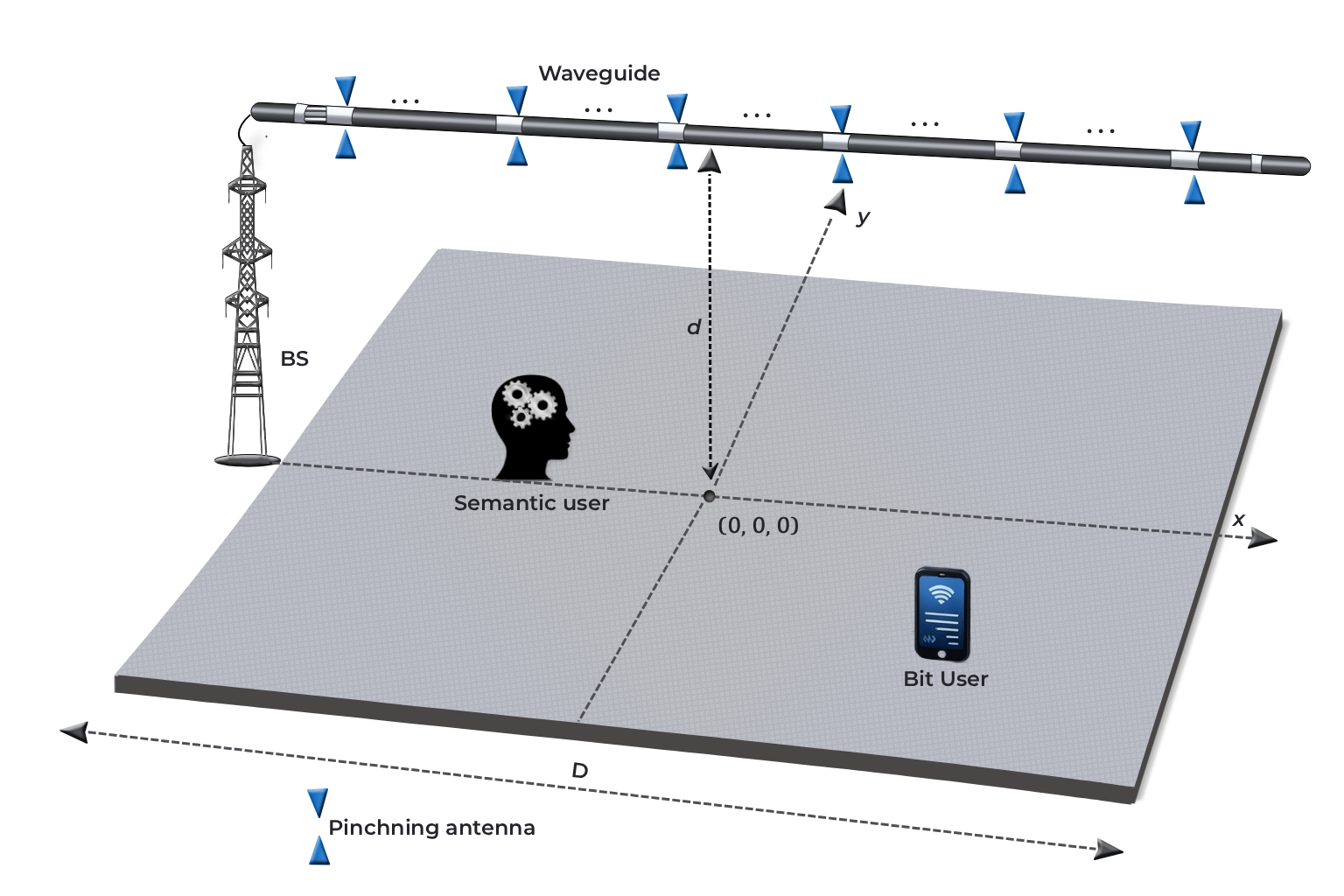}
\caption{An illustration of single-waveguide PASS serving heterogeneous semantic and bit users.}
\vspace{-9mm}
\label{fig1}
\end{figure}

\vspace{-0.0556em}
We consider a downlink NOMA-assisted setup where a base station (BS) is simultaneously serving a semantic user (S) and a bit user (B), via a waveguide mounted at height $d$. The BS is assumed to be directly connected to the waveguide feed point. As shown in Fig.~\ref{fig1}, the users are randomly located in a square region on the Cartesian plane with side length $D$. The waveguide is equipped with $N$ pinching antennas along its length, such that the adjacent antennas are $\Delta\!\geq\!\lambda/2$ apart, where $\lambda$ is the free-space wavelength. Let the $n$-th pinching antenna be at $\tilde{\psi}_n^\mathrm{P}\!=\!(\tilde{x}_n^\mathrm{P}, 0, d)$ with $\tilde{x}_n^\mathrm{P}\!\! \in\![-D / 2, D / 2]$, $n \in N$, and collect their $x$-coordinates in vector $\textbf{x}^\text{P}\!\! =\!\![\tilde{x}_1^\mathrm{P}, \ldots, \tilde{x}_N^\mathrm{P}]$. The single-antenna users are at locations $\phi_m\!=\!\left(x_m, y_m, 0\right)$, where $m\! \in\!\{\mathrm{S},\mathrm{B}\}$, and $x_m$ and $y_m$ are the coordinates in the $xy$-plane.
The free-space channel from the $n$-th antenna to User $m$ is given by the spherical wave model \cite{zhang2022beam} as
\begin{equation}
h_{n,m}\!=\!\frac{\sqrt{\eta} e^{-j \frac{2 \pi}{\lambda}|\phi_m-\tilde{\psi}_n^{\mathrm{P}}|}}{|\phi_m-\tilde{\psi}_n^{\mathrm{P}}|},
\label{channel}
\vspace{-0.68em}
\end{equation}
here $\eta\!=\!\frac{\lambda^2}{16\pi^2}$ is the path loss at 1 \!m reference distance, $|\cdot|$ denotes the Euclidean norm, and $j$ is the imaginary unit of a complex number.

Since all the pinching antennas are on the same waveguide and driven by a single RF chain, the BS must superimpose the signals before transmission as
\begin{equation}
s\!=\!\sqrt{\alpha_\text{S}} s_\text{S}+\sqrt{\alpha_\text{B}} s_\text{B},
\label{superimposed}
\end{equation}
where $s_\text{S}$ and $s_\text{B}$ are the signals intended for semantic and bit users, respectively, with their power allocation coefficients $\alpha_\text{S}$ and $\alpha_\text{B}$, such that $\alpha_\text{S}\!+\!\alpha_\text{B}\!\!=\!\!1$. However, the transmitted signal also includes an additional phase shift $\theta_n$ due to its propagation inside the dielectric waveguide, which lowers its phase velocity relative to free-space. This is captured by the shortened guided wavelength $\lambda_g\!=\!\frac{\lambda}{\eta_\text{eff}}$, where $\eta_\text{eff}$ is the effective refractive index of the dielectric waveguide. Consequently, the transmitted signal vector from each antenna with uniform power distribution \cite{ding2025analytical} can be written as
\begin{equation}
\mathbf{s}\!=\!\sqrt{\frac{P_{\max }}{N}}\left[e^{-j \theta_1}, \cdots, e^{-j \theta_N}\right]^{\mathrm{T}} s,
\label{tx}
\end{equation}
where $P_{\max }$ is the transmit power of BS, $\theta_n\!=\!2 \pi \frac{|\psi_0^{\mathrm{P}}-\tilde{\psi}_n^{\mathrm{P}}|}{\lambda_g}$ is the phase shift at the $n$-th pinching antenna, $[\cdot]^{\mathrm{T}}$ denotes the transpose operation, and $\psi_0^{\mathrm{P}}$ is the feed point to waveguide. The received signal at User $m$ is represented as
\begin{equation}
y_m\!=\!\mathbf{h}_m^T \mathbf{s}+w_m,
\label{received}
\end{equation}
here $w_m$ denotes the additive white Gaussian noise with variance $\sigma^2$, and
\begin{equation}
\mathbf{h}_m\!=\!\left[\frac{\sqrt{\eta} e^{-j \frac{2 \pi}{\lambda}|\phi_m-\tilde{\psi}_1^{\mathrm{P}}|}}{|\phi_m-\tilde{\psi}_1^{\mathrm{P}}|} \; \cdots \; \frac{\sqrt{\eta} e^{-j \frac{2 \pi}{\lambda}|\phi_m-\tilde{\psi}_N^{\mathrm{P}}|}}{|\phi_m-\tilde{\psi}_N^{\mathrm{P}}|}\right]^T\!\!\!\!\!.
\label{channelvector}
\end{equation}
The principle of bit-to-semantic decoding is adopted in the NOMA-assisted PASS. \color{blue} The fixed bit-to-semantic decoding order is adopted due to the heterogeneous receiver structure. The semantic receiver is assumed to include the conventional bit decoder and can therefore decode the bit stream before applying SIC. However, the bit receiver is not assumed to have the trained semantic decoder required for semantic-symbol reconstruction. Hence, the bit receiver directly decodes the bit stream while treating the semantic stream as interference. \color{black} Therefore, the data rate of User B can be formulated as
\begin{equation}
R_\text{B}^\text{P}\left(\textbf{x}^\text{P}, \alpha_\text{S}\right)\!=\!\log _2\left(1+\frac{(1-\alpha_\text{S}) P_{\max } |g_\text{B}|^2}{\alpha_\text{S} P_{\max } |g_\text{B}|^2+\sigma^2}\right),
\label{bitrate}
\end{equation}
where $g_\text{B}\!\!\!=\!\!\!\sum\limits_{n \in N}\frac{\sqrt{\eta} e^{-j \frac{2 \pi}{\lambda}|\phi_\text{B}-\tilde{\psi}_n^{\mathrm{P}}|}}{|\phi_\text{B}-\tilde{\psi}_n^{\mathrm{P}}|}  e^{-j \theta_n}$.
At User S, SIC removes the achievable rate of User B given by
\begin{equation}
R_{\text{B}\rightarrow \text{S}}^\text{P}\left(\textbf{x}^\text{P}, \alpha_\text{S}\right)\!=\!\log _2\left(1+\frac{(1-\alpha_\text{S}) P_{\max } |g_\text{S}|^2}{\alpha_\text{S} P_{\max } |g_\text{S}|^2+\sigma^2}\right),
\label{SICeqrate}
\end{equation}
where $g_\text{S}\!\!\!=\!\!\!\sum\limits_{n \in N}\frac{\sqrt{\eta} e^{-j \frac{2 \pi}{\lambda}|\phi_\text{S}-\tilde{\psi}_n^{\mathrm{P}}|}}{|\phi_\text{S}-\tilde{\psi}_n^{\mathrm{P}}|}  e^{-j \theta_n}$. It should be noted that due to dynamic control of the antennas locations, a sufficiently strong LoS is created for User S, while satisfying the specified minimum bit-rate requirement $R_{\text{B} }^{\text{min}}$. Subsequently, User S decodes its signal in an interference-free manner as
\begin{equation}
R_\text{S}^\text{P}\left(\textbf{x}^\text{P}, \alpha_\text{S}\right)\!=\!\frac{I}{K L} \epsilon_K(\gamma_\text{S})
\label{semanticrate},
\end{equation}
where $\gamma_\text{S}\!=\!\frac{\alpha_\text{S} P_{\max } |g_\text{S}|^2}{\sigma^2}$. In practice, the generalized logistic approximation of $\epsilon_K(\gamma_\text{S})$ is adopted. Specifically, for each $K$ the DeepSC tool is run over a grid of $\gamma_\text{S}$ values to obtain empirical $\epsilon_K(\gamma_\text{S})$ samples. Running the DeepSC with varying values of $K$ and $\gamma_\text{S}$, $\epsilon_K(\gamma_\text{S})$ was found to be monotonically non-decreasing with $\gamma_\text{S}$ \cite{yan2022resource}. Moreover, its gradient change increases first with $\epsilon_K(\gamma_\text{S})$ and then decreases. This pattern suggests that the fitted curve for $\epsilon_K(\gamma_\text{S})$ should look like the sigmoid curve bounded in [0, 1]. Authors in \cite{mu2022heterogeneous} deployed the data-regression method to tractably approximate its values by following the mean square error criterion to express with a generalized logistic function as
\begin{equation}
\epsilon_K(\gamma_\text{S}) \overset{\triangle}{=} A_{K,1} + 
\frac{A_{K,2} - A_{K,1}}{1 + e^{-(C_{K,1} \gamma_\text{S} + C_{K,2})}}.
\label{logisticapprox}
\end{equation}

\color{blue}
The proposed PASS-NOMA resource-allocation framework is not restricted to one fixed sigmoid curve, but it can be applied with any fitted monotonic differentiable semantic-performance function. However, for different DeepSC scenarios (e.g., the existence of co-channel interference due to imperfect SIC), other sophisticated approximation methods might be required and the form of $\epsilon_K\left(\gamma_\text{S} , \boldsymbol{\theta}_K\right)$ will be different. Full end-to-end semantic reconstruction experiments over multiple datasets, neural architectures, and task-oriented metrics are beyond the scope of the present work, and would require training/testing different semantic transceivers and re-fitting the corresponding performance functions.

\color{black}
\vspace{-1em}
\section{Single-Waveguide Problem Formulation}
Our objective is to maximize the SE for User S while guaranteeing the QoS for User B, and invoking the bit-to-semantic decoding order. An optimization problem is formulated so that the rates for both users depend jointly on $\textbf{x}^\text{P}$ and $\alpha_\text{S}$, as given by:
\begin{align}
\mathbf{(P0)}:\max _{\textbf{x}^\text{P}, \alpha_\text{S}} \quad & R_\text{S}^\text{P}\left(\textbf{x}^\text{P}, \alpha_\text{S}\right), \label{originalobjfunc}\\
\text { s.t. } \quad & \left|\tilde{x}_n^\text{P}-\tilde{x}_{n-1}^\text{P}\right| \geq \Delta, \! \forall n \in\{2, \ldots, N\}, \label{antennaspacing}\\
& R_\text{B}^\text{P}\left(\textbf{x}^\text{P}, \alpha_\text{S}\right) \geq R_{\text{B} }^{\text{min}}, \label{bituserQoS}\\
& R_{\text{B}\rightarrow \text{S}}^\text{P}\left(\textbf{x}^\text{P}, \alpha_\text{S}\right) \geq R_{\text{B} }^{\text{min}}, \label{SIC}\\
& 0<\alpha_\text{S}<\alpha_\text{B}. \label{feasiblePA}
\end{align}
Constraint \eqref{antennaspacing} enforces the minimum adjacent antennas spacing to prevent inter-channel coupling, while \eqref{bituserQoS} and \eqref{SIC} respectively ensure bit-user QoS and SIC feasibility under the bit-to-semantic decoding order prescribed by \eqref{feasiblePA}. \color{blue} The bit stream is assigned the higher power level because it corresponds to conventional data transmission with a stringent QoS requirement. The semantic stream is evaluated through a semantic similarity metric and is decoded by the semantic receiver after removing the bit stream. \color{black} The formulated maximization problem is non-convex because $\epsilon_K(\gamma_\text{S})$ fails to satisfy concavity in $\gamma_\text{S}$.

\color{blue} The above SIC feasibility constraint follows the standard NOMA decoding condition under ideal CSI, capacity-approaching coding, and perfect SIC. In practical systems, decoding errors and imperfect interference cancellation causes residual interference and decoding error propagation. A practical reliability-aware design can account for this by introducing an SIC reliability margin as \(R_{\text{B} \rightarrow \text{S}} \geq R_\text{B}^{\min }+\Delta_{\mathrm{SIC}}\), where \(\Delta_{\mathrm{SIC}}\geq 0\) is a prescribed rate margin selected according to practical reliability requirements. Similarly, imperfect SIC can be modelled by a residual interference coefficient \(\rho_{\mathrm{SIC}} \in [0,1]\), where \(\rho_{\mathrm{SIC}}=0\) corresponds to perfect SIC and \(\rho_{\mathrm{SIC}}>0\) represents residual interference after cancellation. In this work, we focus on the ideal SIC case, while imperfect-SIC design is left for future investigation.

\color{black}
\vspace{-1em}
\subsection{Solution Method}
We adopt AO to solve the non-convex problem. \color{blue} At each AO iteration, the algorithm updates one optimization variable while treating the others as fixed. The algorithm is designed to ensure a feasible block-wise solution with monotonic objective function improvement \cite{1-11}. \color{black} For brevity, the AO-based solution approach is summarized in Algorithm 1.
\begin{algorithm}[]
\caption{AO Algorithm for users power allocation and pinching antennas position in single-waveguide PASS}
\begin{algorithmic}[1] 
    \State \textbf{Initialization:} System parameters $(P_{\max},\sigma^2,R_{\text{B} }^{\text{min}},\Delta,d,f_c,\eta_\text{eff})$, SC parameters $(A_{K,1},A_{K,2},C_{K,1},C_{K,2},K,I,L)$, users and BS geometry, iteration index $t \gets 0$, maximum iteration number, initial antenna positions $\textbf{x}^{\text{P}(0)}$.
    \State Compute $h_{\text{S}}$ and $h_{\text{B}}$ via spherical wave model.
    \Repeat
    \State \textbf{Power Allocation Update:}
    \State Set $\tau=2^{R_{\text{B} }^{\text{min}}}-1$.
    \State Compute upper bounds $\alpha_\text{S}=\frac{P_{\max } h_{\text{B}}-\tau \sigma^2}{P_{\max }h_{\text{B}}(1+\tau)}$, and $\alpha_\text{S-SIC}=\frac{P_{\max } h_{\text{S}}-\tau \sigma^2}{P_{\max }h_{\text{S}}(1+\tau)}$.
    \State Update $\alpha_\text{S}^{(t)} = \max \left\{0, \min \left\{\alpha_\text{S}, \alpha_\text{S-SIC}, 0.5\right\}\right\}$.
    \If{$\alpha_\text{S}^{(t)}=0$}
        \State Infeasible antenna positions.
        \Else
        \State $\alpha_\text{S}^*=\alpha_\text{S}^{(t)}$, and $\gamma_\text{S}^* = \tfrac{\alpha_\text{S}^*P_{\max }h_{\text{S}}}{\sigma^2}$.
    \EndIf
    \State \textbf{Position Update:} 
    \State Adjust $\textbf{x}^{\text{P}(t)}$ via bisection search towards the semantic user to increase $h_{\text{S}}$, while enforcing $|\tilde{x}_n^\text{P}-\tilde{x}_{n-1}^\text{P}|\geq\Delta$.
    \State Set $c_{\text{left}} \gets x_{\text{S}}$, $c_{\text{right}} \gets x_{\text{B}}$, $c_{\text{mid}} \gets (c_{\text{left}} + c_{\text{right}})/2$, tolerance $\varepsilon$, phase fine-tuning step $\tilde{\Delta}$, and precision constants $\delta_\text{S}$, $\delta_\text{B}$.
     \State Compute updated $h_{\text{S}}$ and $h_{\text{B}}$.
     \If{$R_\text{B}^\text{P}(c_\text{mid}) \geq R_{\text{B} }^{\text{min}}$ \textbf{and}      $R_{\text{B}\rightarrow\text{S}}^\text{P}(c_\text{mid}) \geq R_{\text{B} }^{\text{min}}$}
     \State $c_{\text{right}} \gets c_{\text{mid}}$
     \Else
     \State
          $c_{\text{left}} \gets c_{\text{mid}}$
    \EndIf
    \State Fine-tune remaining antennas by $\pm \tilde{\Delta}$ to satisfy ajacent spacing and phase errors $\le \delta_\text{S}$, $\delta_\text{B}$.
    \State $t \gets t+1$
    \Until{convergence or maximum iteration number reached.}
\State \Return $\alpha_\text{S}^*$, $\textbf{x}^\text{P*}$ and $R_\text{S}^\text{P*}$.
\end{algorithmic}
\end{algorithm}

\vspace{-1em}
\subsection{Power Allocation Subproblem} \label{singleWGuserspower}
In this subproblem, $\textbf{x}^\text{P}$ is assumed to be fixed and feasible, which simplifies $\mathbf{(P0)}$ as:
\begin{align}
\max _{\alpha_\text{S}} \quad & R_\text{S}^\text{P}\left( \alpha_\text{S}\right), \label{PAobjfunc}\\
\text { s.t. } \quad
& R_\text{B}^\text{P}\left(\alpha_\text{S}\right) \geq R_{\text{B} }^{\text{min}}, \label{PAbituserQoS}\\
& R_{\text{B}\rightarrow \text{S}}^\text{P}\left(\alpha_\text{S}\right) \geq R_{\text{B} }^{\text{min}}, \label{PASIC}\\
& 0<\alpha_\text{S}<\alpha_\text{B}. \label{subfeasiblePA}
\end{align}
It is important to consider that the above simplified problem is non-convex due to the dependence on $\epsilon_K$. However, the one-dimensional power allocation subproblem satisfies the ``time-sharing" criterion \cite{yu2006dual}, allowing the Lagrangian functions to approximate the optimal solution with zero duality gap.

Users power allocation is decided on the basis of active constraints for the bounded objective function. From \eqref{PAbituserQoS}, algebraic manipulations yield the closed-form solution as
\begin{equation}
\alpha_\text{S} \leq \frac{P_{\max } h_{\text{B}}-\tau \sigma^2}{P_{\max }h_{\text{B}}(1+\tau)},
\label{alphas}
\end{equation}
where $\tau=2^{R_{\text{B} }^{\text{min}}}-1$, and $P_{\max } h_{\text{B}}-\tau \sigma^2 \geq 0$ for feasibility at the given $\textbf{x}^\text{P}$. Likewise, from \eqref{PASIC}, the SIC decodability constraint results in the following closed-form upper bound $\alpha_\text{S-SIC}$ on the semantic-user power allocation coefficient.
\begin{equation}
\alpha_\text{S-SIC} \leq \frac{P_{\max } h_{\text{S}}-\tau \sigma^2}{P_{\max }h_{\text{S}}(1+\tau)},
\label{alphasic}
\end{equation}
Following the proof in \cite{fu2025power}, the optimal power coefficient is obtained when the closed-form solutions hold as equalities. The optimized power allocation coefficient $\alpha_\text{S}^*$ with the upper bound value can therefore be written as
\begin{equation}
\alpha_\text{S}^*=\max \left\{0, \min \left\{\alpha_\text{S}, \alpha_\text{S-SIC}, 0.5\right\}\right\}.
\label{alphas*}
\end{equation}

\vspace{-1em}
\subsection{Antennas Position Subproblem}  \label{singleWGpinchingantennas}
In this subproblem, our aim is to strategically determine the deployment of pinching antennas based on a fixed $\alpha_\text{S}^*$ value. \color{blue} The pinching antenna position affects the channel through two coupled mechanisms, namely the distance-dependent spherical-wave free-space component and the guided-wave phase component along the dielectric waveguide. \color{black} Consequently, determining the optimal pinching antennas position vector $\textbf{x}^\text{P*}\!\! =\!\![\tilde{x}_1^\mathrm{P*}, \ldots, \tilde{x}_N^\mathrm{P*}]$ that maximizes the semantic SE in \eqref{originalobjfunc} is of great importance. Moreover, to cope with the phase shifts due to the propagation along the waveguide, it is necessary to fine-tune the pinching antennas deployment to ensure their phase alignment for the signal through free-space following waveguide propagation. Therefore, this subproblem involves two coordinated steps, namely large-scale antenna placement, and fine-scale phase alignment.

Let us assume that the pinching antennas are placed sequentially along the x-axis on the waveguide, with adjacent ones satisfying the minimum-spacing condition. Based on this, the antenna position subproblem is formulated as:
\begin{align}
\max _{\textbf{x}^\text{P}} \quad & R_\text{S}^\text{P}\left( \textbf{x}^\text{P}\right), \label{APobjfunc}\\
\text { s.t. } \quad & \left|\tilde{x}_n^\mathrm{P}-\tilde{x}_{n-1}^\mathrm{P}\right| \geq \Delta, \! \forall n \in\{2, \ldots, N\}, \label{APspacing}\\
& R_\text{B}^\text{P}\left(\textbf{x}^\text{P}\right) \geq R_{\text{B} }^{\text{min}}, \label{APbituserQoS}\\
& R_{\text{B}\rightarrow \text{S}}^\text{P}\left(\textbf{x}^\text{P}\right) \geq R_{\text{B} }^{\text{min}}, \label{APSIC}\\
& \left|\phi_{\text{S}, n}-\phi_{\text{S}, n-1} \pm 2\pi l\right| \leq \delta_\text{S}, \! \forall n \in\{2, \ldots, N\}, \label{phaseS}\\
& \left|\phi_{\text{B}, n}-\phi_{\text{B}, n-1} \pm 2\pi l\right| \leq \delta_\text{B}, \! \forall n \in\{2, \ldots, N\} \label{phaseSB}
\end{align}
where $\phi_{m,n}\!\!=\!\!2\pi \left( \frac{|\phi_m - \tilde{\psi}_n^{\mathrm{P}}|}{\lambda} - \frac{|\psi_0^{\mathrm{P}} - \tilde{\psi}_n^{\mathrm{P}}|}{\lambda_g} \right)$, such that $m\! \in\!\{\mathrm{S},\mathrm{B}\}$ and includes the free-space and waveguide propagation distance terms.

Due to the non-convex objective function in \eqref{APobjfunc} and the coupled influence of antenna positions on both the LoS gains and the phases seen by the two users, direct analysis is extremely complex. Therefore, we proceed with an iterative based solution for a manageable analysis. Initially, we iteratively relocate the pinching antennas on the waveguide to form the large-scale channel and enhance the effective channel gain of the semantic user while maintaining the bit-user QoS. This is implemented via a one-dimensional bisection search such that the first antenna is placed where the User S experiences the best LoS channel gain, with the remaining evenly placed and satisfying the minimum-spacing constraint. Subsequently, small phase adjustments are performed to maximize constructive interference and thereby enhance the composite channel gain of the semantic user. 
Constraints \eqref{phaseS} and \eqref{phaseSB} ensure phase alignment for the semantic and bit users, respectively. Here, $l$ denotes an arbitrary integer that accounts for the $2 \pi$ phase periodicity, while $\delta_\text{S}$ and $\delta_\text{B}$ represent the predefined positive phase-precision constants for these users. \color{blue} For any User $m$, the adjacent antennas phase difference is computed as $\Delta \phi_{m, n}\!\!\!=\!\!\!\phi_{m, n}-\phi_{m, n-1}$. Since phase is periodic, the nearest integer multiple of $2 \pi$ is removed. Equivalently, the residual wrapped phase error is computed as $e_{m, n}\!=\!\Delta \phi_{m, n}-2 \pi l^{\star}$, where $\displaystyle l^{\star}\!=\! \arg\min_{l \in \mathbb{Z}}\left|\Delta \phi_{m, n}-2 \pi l\right|$. The phase-precision condition is satisfied if $\left|e_{m, n}\right| \leq \delta_m$.

Notably, the phase-precision conditions are not assumed to be always jointly satisfiable for every channel realization. Instead, they are used as feasibility and phase-refinement criteria within the position update step. Specifically, after the large-scale bisection-based antennas placement, the phase fine-tuning step searches around the candidate positions to reduce the residual phase mismatch. A fine-tuned candidate is accepted only if it satisfies the minimum-spacing constraint, the semantic-user and bit-user phase-precision conditions, the bit-user QoS constraint, and the SIC feasibility constraint. Otherwise, the candidate is discarded and the previous feasible pinching antennas placement is retained. If no feasible placement exists for a given channel realization, that realization is treated as infeasible under the considered constraints. Therefore, the accepted pinching antennas-position sequence is non-decreasing by construction, even though the search used to generate candidate positions is heuristic.

\color{black}
\vspace{-1.5em}
\subsection{Complexity and Convergence Analysis}
For each channel realization, the proposed AO algorithm alternates between a power allocation update and a position update. In the first step, obtaining $\alpha_\text{S}^*$ requires a constant computational complexity $\mathcal{O}(1)$, where $\mathcal{O}$ denotes the big-O notation. In the second step, each bisection search requires summing the contributions of all $N$ pinching antennas, thus costing $\mathcal{O}(N)$. The number of bisection iterations required to shrink the initial search interval of length $M$ to a convergence tolerance $\varepsilon$ is on the order of $\mathcal{O}(\log _2(M / \varepsilon))$. Subsequently, the stage of improving phase alignment at the semantic user is linear in $N$ and does not affect the overall order. Therefore, the total complexity of Algorithm 1 is $\mathcal{O}(N\log _2(M / \varepsilon))$.

In each AO iteration, the closed-form update of $\alpha_\text{S}$ maximizes the semantic SE within the feasible power interval for a fixed $\textbf{x}^\text{P}$. Conversely, for a fixed $\alpha_\text{S}$, the position update step ensures that the current semantic SE does not decrease while maintaining all feasibility constraints. Consequently, the sequence of semantic rates generated is monotonically non-decreasing and bounded. Therefore, the proposed algorithm converges to a finite limit of $R_\text{S}^\text{P}$.

\vspace{-0.759975em}
\section{Multi-Waveguide System Model}
\begin{figure}[!t]
\centering
 \includegraphics[trim={10cm 8cm 10cm 2cm},clip,width=1\columnwidth]{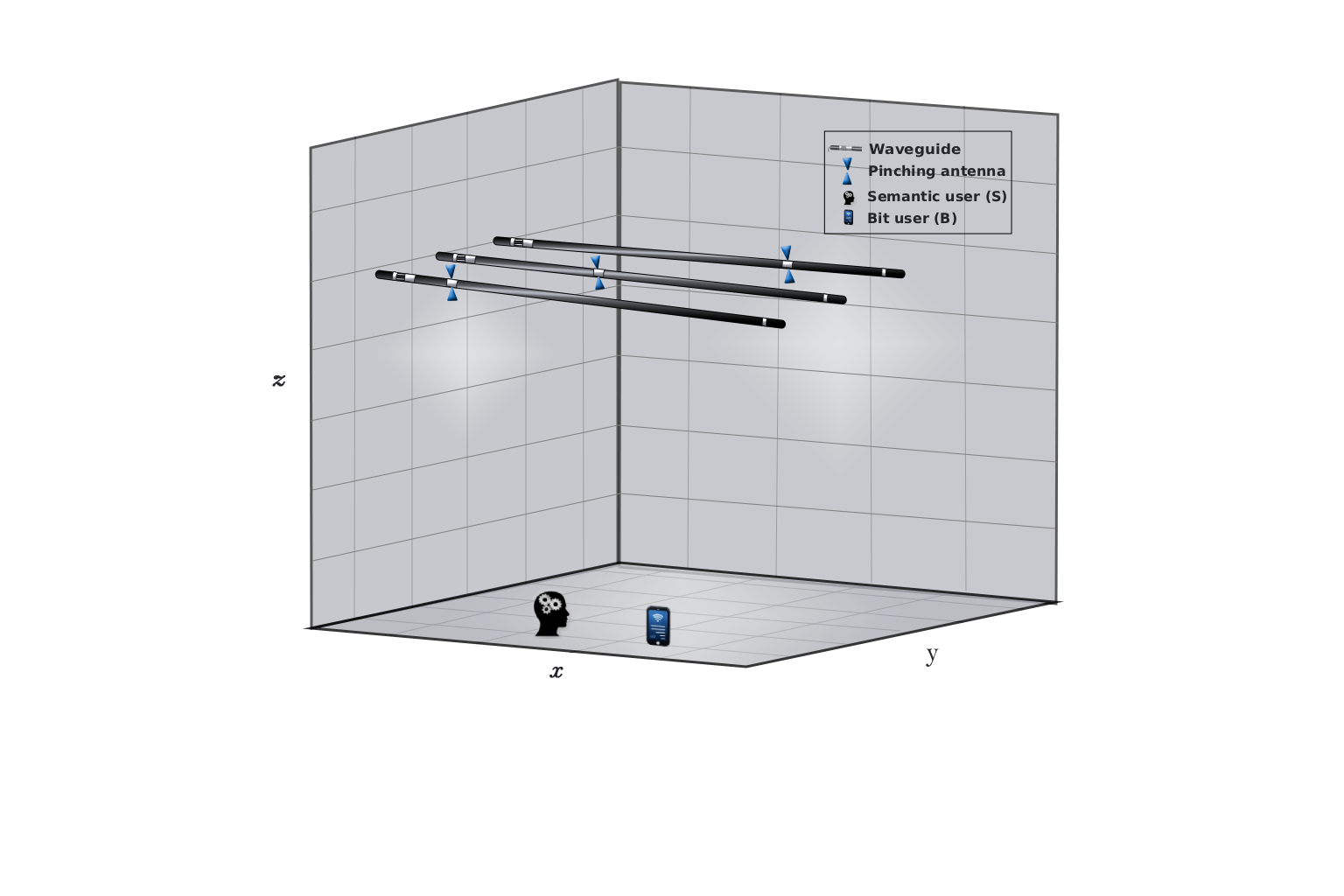}
\caption{Geometry of the multi-waveguide PASS.}
\vspace{-9mm}
\label{multiwg}
\end{figure}

\vspace{-0.0556em}
We now extend the single-waveguide PASS architecture to multiple waveguides such that each waveguide carries one pinching antenna. \color{blue}Moreover, the two-user framework can be extended to multi-user NOMA with multiple clusters, as detailed in Appendix. \color{black}Let us consider that the BS at height $d$ is equipped with a set of $K_\text{wg}$ waveguides and collectively serving S and B users within the same squared region as in the single-waveguide setup. Each waveguide is placed at a distinct lateral offset $y^{(k)}$ such that the inter-waveguide coupling is avoided by ensuring that the pinching antennas on different waveguides are sufficiently apart. The proposed multi-waveguide setup allows each waveguide to transmit the superimposed signal of both users. A passive power splitting network distributes that signal into $K_\text{wg}$ waveguides with fractions of $P_{\max }$. The objective is to exploit the coherent combining across waveguides for the heterogeneous users framework. Let the $k$-th waveguide carries its only pinching antenna at $\tilde{\psi}_k^\mathrm{P}\!\!=\!\!(\tilde{x}_k^\mathrm{P}, y^{(k)}, d)$, where $k\!=\!1,...,K_\text{wg}$ and $\tilde{x}_k^\mathrm{P}$ is the longitudinal position on waveguide for stacking into the variable $\bar{\textbf{x}}\!=\![\tilde{x}_1^\mathrm{P}, \ldots, \tilde{x}_{K_\text{wg}}^\mathrm{P}]$. Geometry of the proposed downlink multi-waveguide setup is as in Fig.~\ref{multiwg}.

The complex channel gain seen by User $m\! \in\!\!\{\mathrm{S},\mathrm{B}\}$ at $\phi_m$ from the $k$-th waveguide is $\tilde{h}_{k, m}\!=\!h_{k, m}^{\mathrm{FS}} e^{-j \theta_k}$, where
\begin{equation}
h_{k, m}^{\mathrm{FS}}=\frac{\sqrt{\eta} e^{-j \frac{2 \pi}{\lambda}|\phi_m-\tilde{\psi}_k^{\mathrm{P}}|}}{|\phi_m-\tilde{\psi}_k^{\mathrm{P}}|},
\end{equation}
and $\theta_k\!=\!2 \pi \frac{|\psi_{0,k}^{\mathrm{P}}-\tilde{\psi}_k^{\mathrm{P}}|}{\lambda_g}$ is the guided propagation phase along the $k$-th waveguide from its feed $\psi_{0,k}^{\mathrm{P}}$. The effective channels
can be obtained by coherently combining the gains from all antennas as
\begin{equation}
h_{m}^{\mathrm{eff}}\!=\!\sum_{k=1}^{K_\text{wg}} \sqrt{\beta_k} \tilde{h}_{k, m},
\label{3wgeffectivechannel}
\end{equation}
where $\beta_k$ is the power allocated to the $k$-th waveguide such that $\boldsymbol{\beta}\!=\!\left[\beta_1,...,\beta_{K_\text{wg}}\right]$. The received signal at User $m$ is
\begin{equation}
r_m\!=\!\sqrt{P_{\max }}h_{m}^{\mathrm{eff}}s+w_m,
\label{3wgreceived}
\end{equation}

\vspace{-1em}
\section{Multi-Waveguide Problem Formulation}
We aim to maximize the semantic SE by formulating a joint optimization problem in which we incorporate an additional optimization variable $\boldsymbol{\beta}$ for the waveguide power allocation together with $\bar{\textbf{x}}$ and $\alpha_\text{S}$, given as follows:
\begin{align}
\mathbf{(P1)}:\max _{\bar{\textbf{x}}, \alpha_\text{S}, \boldsymbol{\beta}} \quad & \bar{R}_\text{S}^\text{P}\left(\bar{\textbf{x}}, \alpha_\text{S}, \boldsymbol{\beta}\right),\label{3wgoriginalobjfunc}\\
\text { s.t. } \quad &  \bar{R}_\text{B}^\text{P}\left(\bar{\textbf{x}}, \alpha_\text{S}, \boldsymbol{\beta}\right) \geq R_{\text{B} }^{\text{min}}, \label{3wgQoS}\\
& \bar{R}_{\text{B}\rightarrow \text{S}}^\text{P}\left(\bar{\textbf{x}}, \alpha_\text{S}, \boldsymbol{\beta}\right) \geq R_{\text{B} }^{\text{min}}, \label{3wgSIC}\\
& 0<\alpha_\text{S}<\alpha_\text{B}, \label{3wgfeasiblePA}\\
& \beta_k \geq 0, \label{waveguidePA}\\
& \sum_{k=1}^{K_\text{wg}} \beta_k=1, \label{waveguidetot}\\
& \bar{\textbf{x}} \in \mathcal{X},
\end{align}
where $\mathcal{X}$ denotes the feasible deployment region along each waveguide to avoid inter-waveguide coupling. Constraints \eqref{3wgQoS} to \eqref{3wgfeasiblePA} respectively ensure bit-user QoS, SIC feasibility, and adopted decoding order. Constraints \eqref{3wgfeasiblePA} and \eqref{waveguidetot} imply that the power fractions allocated to waveguides are within the feasible values. The optimization problem is highly involved due to the strongly coupled optimization variables and the non-concave objective, rendering it as non-convex.

\vspace{-1.3em}
\subsection{Solution Method}
We decouple the problem via AO into three subproblems for the multi-variables update. \color{blue} An update in one block is accepted only if the complete set of relevant system constraints remains satisfied. If a candidate update violates any feasibility condition or fails to preserve the objective value, the previous feasible point is retained. This accept/reject mechanism prevents infeasible transitions between AO iterations. \color{black} Users power allocation and pinching antenna position updates are respectively obtained by following a similar approach as in Subsections \ref{singleWGuserspower} and \ref{singleWGpinchingantennas}, with the exception of no requirement for phase fine-tuning in the multi-waveguide system model as each antenna is mounted on a distinct waveguide with sufficient lateral spacing. \color{blue} For the power-allocation variable/update, the optimal $\alpha_\text{S}$ is obtained in closed form over the feasible interval induced by the QoS, SIC, and NOMA power-domain constraints, hence, this block is optimally solved. For the pinching-antenna position update, the subproblem remains non-convex; therefore, the proposed position update is not claimed to globally solve the original position subproblem. Instead, it is implemented as a structured feasible-search step in which a candidate position is accepted only if it satisfies the bit-user QoS and SIC constraints and does not reduce the semantic objective; otherwise the previous feasible position is retained.  For the multi-waveguide power-splitting block, the MM surrogate satisfies the standard tightness and lower-bound conditions, i.e., the surrogate is equal to the original function at the current point and lower-bounds it elsewhere. The simplex grid search then provides a grid-optimal solution over the discretized feasible simplex. Therefore, each accepted AO/MM update generates a non-decreasing semantic-SE sequence. \color{black} Next, we will elaborate on the waveguides power allocation.

\vspace{-1.3em}
\subsection{Waveguides Power Allocation Subproblem}
For given $\bar{\textbf{x}}\!=\!\bar{\textbf{x}}^{(t)}$ and $\alpha_\text{S}\!=\!\alpha_\text{S}^{(t)}$ parameters, the waveguides power allocation subproblem can be written as:
\begin{align}
\max_{\boldsymbol{\beta}} \quad &
\bar{R}_\text{S}^\text{P}(\bar{\mathbf{x}}^{(t)}, \alpha_\text{S}^{(t)}, \boldsymbol{\beta}),
\label{3wgobjfuncpowerallocation}\\
\text{s.t.} \quad &
\bar{R}_\text{B}^\text{P}(\bar{\mathbf{x}}^{(t)}, \alpha_\text{S}^{(t)}, \boldsymbol{\beta})
\geq R_{\text{B}}^{\text{min}},
\label{3wgQoSpowerallocation}\\
&
\bar{R}_{\text{B}\rightarrow \text{S}}^\text{P}(\bar{\mathbf{x}}^{(t)}, \alpha_\text{S}^{(t)}, \boldsymbol{\beta})
\geq R_{\text{B}}^{\text{min}},
\label{3wgSICpowerallocation}\\
& \boldsymbol{\beta} \in \Delta_{K_\text{wg}},
\end{align}

where $\Delta_{K_\text{wg}}$ is the feasible set of power allocations across the waveguides and defined as
\begin{equation}
\Delta_{K_\text{wg}}
\!=\!  \{\boldsymbol{\beta} \in \mathbb{R}^{K_\text{wg}} :
\beta_k \ge 0,\ \textstyle\sum\limits_{k=1}^{K_\text{wg}} \beta_k = 1  \}.
\end{equation}
Here, $\mathbb{R}^{K_\text{wg}}$ denotes the $K_\text{wg}$-th dimensional real vector.

The objective remains to maximize the semantic SE such that for each $\boldsymbol{\beta}$ update, the non-decreasing trend of $\epsilon_K(\gamma_\text{S})$ should not be violated. Based on the adopted logistic approximation, the waveguides power allocation can be interpreted as a monotonic maximization problem \cite{zhang2013monotonic}, where improvement in $\gamma_\text{S}$ cannot reduce the objective. In particular, the optimization problem is still non-convex and the rate expressions in \eqref{3wgobjfuncpowerallocation} to \eqref{3wgSICpowerallocation} depend on coherent combining via $\sqrt{\beta_k}$, rendering its solution highly challenging.

Therefore, inspired with the monotonic optimization theory, we adopt an MM approach that iteratively maximizes a tight global lower bound surrogate for a guaranteed non-decreasing monotonic objective sequence \cite{sun2016majorization}, \cite{nguyen2017introduction}. For notational convenience, we introduce the auxiliary variable $z_k\!=\!\sqrt{\beta_k}$ and transform the problem as follows.

\vspace{-1.3em}
\subsection{Problem Reformulation and Surrogate Construction}
The coherent combining in \eqref{3wgeffectivechannel} can be written as
\begin{equation}
h_{m}^{\mathrm{eff}}\!=\!\tilde{\mathbf{h}}_m^T \mathbf{z},
\label{3wgcompacteffectivechannel}
\end{equation}
where $\tilde{\mathbf{h}}_m\!\!=\!\![\tilde{h}_{1, m}, \ldots, \tilde{h}_{K_\text{wg}, m}]^T$ and $\mathbf{z}\!=\![z_1, \ldots, z_{K_\text{wg}}]^T$, such that $|\mathbf{z}|^2\!=\!1$. With this, the received signal power taking the quadratic form at User $m$ can be equivalently as
\begin{equation}
q_m(\mathbf{z})\!=\!\mathbf{z}^T \mathbf{Q}_m \mathbf{z},
\label{eqreceivedpowerterm}
\end{equation}
where $\mathbf{Q}_m \triangleq \Re (\tilde{\mathbf{h}}_m^* \tilde{\mathbf{h}}_m^T) \succeq 0$ such that $\Re$ represents the real part of a complex number.

The bit-user QoS and SIC feasibility constraints in \eqref{3wgQoSpowerallocation} and \eqref{3wgSICpowerallocation} are equivalently simplified with the following respective lower bounds as
\begin{equation}
            q_\text{B}(\mathbf{z}) \geq T_\text{B},
\end{equation}
\begin{equation}
            q_\text{S}(\mathbf{z}) \geq T_\text{B},
\end{equation}
where $T_\text{B}\!\!=\!\!\frac{\tau \sigma^2}{P_{\max }(\alpha_\text{B}-\tau \alpha_\text{S})}$, such that $\alpha_\text{B}-\tau \alpha_\text{S}\!\!>\!\!0$ for feasibility.

Based on the above discussion, the maximization problem is reformulated as:
\begin{equation}
\begin{array}{cl}
\max \limits_{\mathbf{z}} & \mathbf{z}^T \mathbf{Q}_\text{S} \mathbf{z} \\
\text { s.t. } & \mathbf{z}^T \mathbf{Q}_\text{B} \mathbf{z} \geq T_\text{B}, \\
& \mathbf{z}^T \mathbf{Q}_\text{S} \mathbf{z} \geq T_\text{B}, \\
& \mathbf{z} \succeq 0,\ |\mathbf{z}|^2=1.
\end{array}
\end{equation}
Solution of the above maximization problem requires constructing a surrogate that lower bounds the objective function. At each iteration, the resulting surrogate is then maximized to obtain a non-decreasing update for the objective function.

Since $\mathbf{Q}_m \succeq 0$, the quadratic term $q_m(\mathbf{z})$ is convex in $\mathbf{z}$, and its first-order Taylor expansion constitutes a global affine lower bound.
\begin{equation}
q_m(\mathbf{z}) \geq \underline{q}_m(\mathbf{z} \mid \mathbf{z}^{(i)})=q_m(\mathbf{z}^{(i)})+\nabla q_m(\mathbf{z}^{(i)})^T(\mathbf{z}-\mathbf{z}^{(i)}),
\end{equation}
where $\underline{q}_m$ is the lower bound surrogate and $\nabla$ denotes the gradient operator. Since $\nabla q_m(\mathbf{z})\!=\!2 \mathbf{Q}_m \mathbf{z}$, this yields
\begin{equation}
\underline{q}_m(\mathbf{z} \mid \mathbf{z}^{(i)})=2(\mathbf{Q}_m \mathbf{z}^{(i)})^T \mathbf{z}-(\mathbf{z}^{(i)})^T \mathbf{Q}_m \mathbf{z}^{(i)}.
\label{surrogate}
\end{equation}
This surrogate satisfies the standard MM conditions of tightness and global lower boundedness, i.e., $\underline{q}_m(\mathbf{z}^{(i)} \mid \mathbf{z}^{(i)})=q_m(\mathbf{z}^{(i)})$ and $\underline{q}_m(\mathbf{z} \mid \mathbf{z}^{(i)}) \leq q_m(\mathbf{z})$ $\forall \mathbf{z}$.

The waveguides power allocation is obtained by solving the surrogate function in \eqref{surrogate} iteratively for the linear constraints over a unit simplex. In the considered setup, it is efficiently solved via a simplex grid search, yielding the optimal solution $\mathbf{z}^{(i+1)}$, followed by the update $\beta_k\!=\!z_k^2$. This yields a feasible power split satisfying the bit-user QoS and SIC constraints. Since each surrogate $\underline{q}_\text{S}(\mathbf{z} \mid \mathbf{z}^{(i)})$ is a tight global lower bound of the true quadratic $q_\text{S}(\mathbf{z})$ at $\mathbf{z}^{(i)}$, and the feasibility is determined conservatively using the same lower bounds, the MM update ensures $R_\text{S}(\boldsymbol{\beta}^{(i+1)}) \geq R_\text{S}(\boldsymbol{\beta}^{(i)})$. The resulting $\beta^{(i+1)}$ is subsequently used to update $\bar{\textbf{x}}$ and  $\alpha_\text{S}$.

\color{blue} In the ideal MM procedure, monotonicity is guaranteed when the surrogate problem is solved exactly, or more generally when the selected update gives a surrogate objective value no smaller than that at the current iterate. In our case, the surrogate function is a tight lower bound of the original quadratic semantic-channel gain at the current iterate. Therefore, if a new grid point gives a surrogate value no smaller than the value at the current iterate, then the semantic-channel gain is also non-decreasing. Furthermore, the semantic SE is a monotonically increasing function of the semantic-user SNR under the adopted semantic similarity approximation, since $A_{K, 2}>A_{K, 1}$ and $C_{K, 1}>0$. Hence, a non-decreasing semantic-user effective channel gain leads to a non-decreasing semantic-user SNR, which in turn leads to a non-decreasing semantic SE.

\color{black}
Since the $\mathbf{Q}_m \mathbf{z}^{(i)}$ product in each MM iteration involves a dense multiplication between a $K_\text{wg} \!\times\! K_\text{wg}$ matrix and a $K_\text{wg}$-th dimensional vector, its computational cost scales as $\mathcal{O}(K_\text{wg}^2)$. Moreover, the surrogate maximization is performed over the simplex constraint \(\sum_{k=1}^{K_\text{wg}} \beta_k \!=\! 1\), offering $K_\text{wg}-1$ degrees of freedom. Therefore, a uniform simplex grid search with step size $r$ requires on the order of $\mathcal{O}((1 / r)^{K_\text{wg}-1})$ candidate evaluations.
\color{blue} In the multi-waveguide case, each channel evaluation during the bisection-based position update requires coherent summation over \(K_{\mathrm{wg}}\) waveguides, and hence costs \(\mathcal{O}(K_{\mathrm{wg}})\). Since the number of bisection iterations required to shrink the initial search interval of length \(M\) to a convergence tolerance \(\varepsilon\) is \(\mathcal{O}(\log_2(M/\varepsilon))\), the position update costs \(\mathcal{O}(K_{\mathrm{wg}}\log_2(M/\varepsilon))\). Combining this with the MM-based waveguide power-splitting update, and considering \(I_{\mathrm{MM}}\) (inner) MM iterations and \(I_{\mathrm{AO}}\) (outer) AO iterations, the overall complexity of the multi-waveguide PASS algorithm becomes
\begin{equation}
\resizebox{\dimexpr\columnwidth-3em\relax}{!}{$
\mathcal{O}\left(
I_{\mathrm{AO}}
\left[
K_{\mathrm{wg}}\log_2\!\left(\frac{M}{\epsilon}\right)
+
I_{\mathrm{MM}}
\left(
K_{\mathrm{wg}}^2
+
K_{\mathrm{wg}}
\left(\frac{1}{r}\right)^{K_{\mathrm{wg}}-1}
\right)
\right]
\right)
$}
\label{eq:complexity}
\end{equation}
This expression shows that the simplex-grid search grows rapidly with \(K_{\mathrm{wg}}\). Therefore, the current grid-based implementation is suitable for the small-dimensional multi-waveguide setting considered in this work.

\color{blue} At each AO iteration, the accepted update of every block is designed not to decrease the semantic SE. Since $\alpha_\text{S}^*\!\!=\!\!\max \left\{0, \min \left\{\alpha_\text{S}, \alpha_\text{S-SIC}, 0.5\right\}\right\}$, $\beta_k \!\geq\! 0$, $\sum_{k=1}^{K_\text{wg}} \beta_k\!=\!1$, and the antenna positions are restricted to a bounded deployment region, the feasible set is closed and bounded. Moreover, the semantic similarity function is bounded. Hence, the sequence of semantic SE values by objective function is monotonically non-decreasing and upper bounded, and therefore admits a finite limiting value under the given accept/reject mechanism.
\color{black}
\vspace{-2.5em}
\section{Simulation Results}
Unless stated otherwise, the simulation parameters used to evaluate the performance of heterogeneous users NOMA with PASS are set as follows: $\sigma^2=-90$\! dBm, the carrier frequency $f_c\!=\!28$\! GHz, $d\!=\!3$\! m, $\eta_\text{eff}=\!1.4$, $\Delta\!=\!\lambda/2$, $\tilde{\Delta}\!=\!\lambda/10$, $R_{\text{B} }^{\text{min}}\!=\!0.5$\! bps/Hz, $D\!\in\!\{20,40\}$\!\! m, and $N\!\!\in\!\!\{3,7\}$, where $\lambda\approx10.7$ mm. At the given $f_c$, due to the extended aperture of the waveguide, which is much larger than $\lambda$, the considered users lie in the near-field region \cite{liu2025pinching}. For SC, the parameters of $K\!=\!5$, $\mu\!=\!40$, $I/L=1$, $A_{K,1}\!=\!0.37$, $A_{K,2}\!=\!0.98$, $C_{K,1}\!=\!0.25$, and $C_{K,2}\!=\!-0.7895$ are adopted as in \cite{mu2023exploiting}. As a benchmark, a NOMA-assisted CAS is considered as in \cite{ding2025flexible, wang2025antenna}, where the BS is equipped with the same number of fixed antennas, centered within the service region at height $d$ and antennas separated by half a wavelength. Unlike PASS, CAS lacks positional flexibility and therefore generally incurs larger BS-to-users path loss. Moreover, the semantic SE results are averaged over $10^5$ realizations.

\begin{figure}[!t]
\centering
\includegraphics[trim={0cm 0cm 0cm 0cm},clip,width=1\columnwidth]{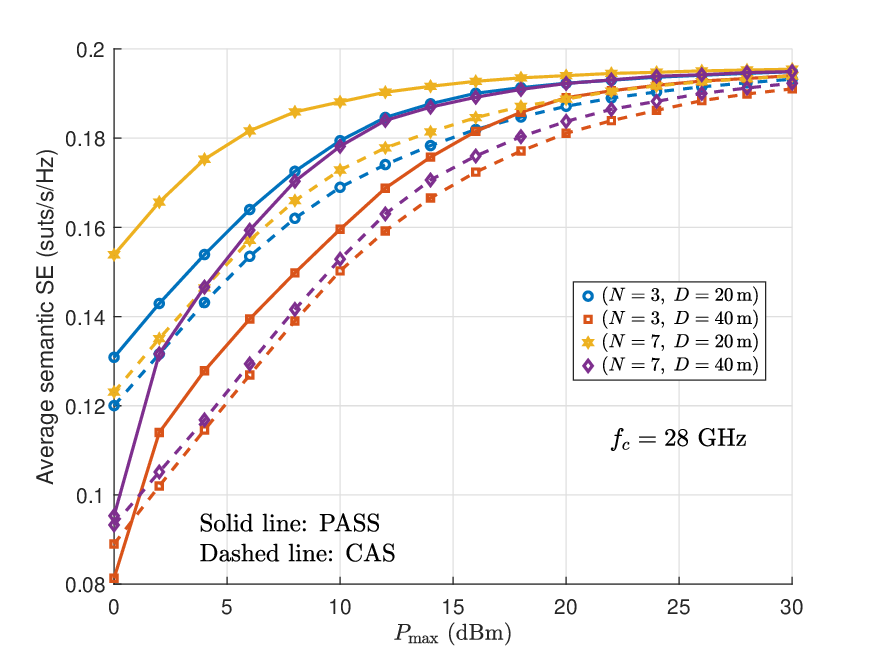}
\caption{Average semantic SE versus $P_{\max }$ for the NOMA assisted single-waveguide PASS and CAS.}
\vspace{-9mm}
\label{fig2}
\end{figure}

Fig.~\ref{fig2} illustrates the average semantic SE versus $P_{\max}$ for the single-waveguide PASS and CAS. As $P_{\max}$ increases, the average semantic SE improves quickly first and then becomes more gradual, consistent with the behavior of SC. In all settings, PASS achieve a higher semantic SE than CAS due to the inherent ability of creating reconfigurable propagation channels via pinching antennas, reinforcing channel disparity between the heterogeneous NOMA users. Increasing the number of antennas further boosts semantic SE due to greater spatial degrees of freedom for an enhanced channel gain. Furthermore, a smaller coverage area offers better semantic SE due to shorter links, while the PASS continuously hold advantage over CAS in all deployment areas. In all cases, the semantic power fraction $\alpha_S$ is selected on the boundary defined by the QoS constraint, and pinching locations satisfy the minimum spacing criteria, turning PASS’s geometric flexibility into clear semantic-rate benefits over CAS.

\begin{figure}[!t]
\centering
\includegraphics[trim={0cm 0cm 0cm 0cm},clip,width=1\columnwidth]{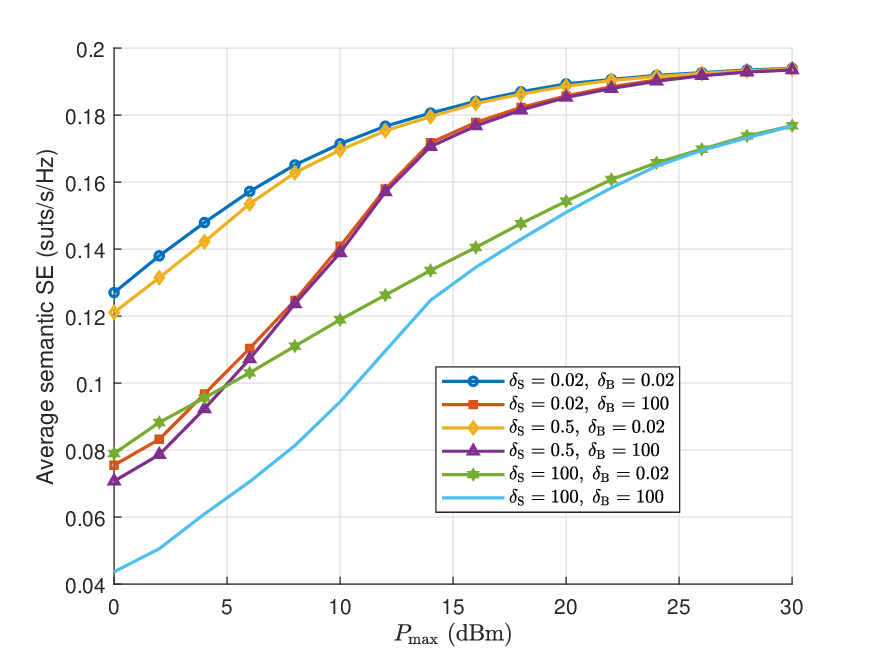}
\caption{Average semantic SE versus $P_{\max }$ under different phase alignments for the single-waveguide PASS with $N\!=\!3$ and $D\!=\!20$\! m.}
\vspace{-9mm}
\label{fig3}
\end{figure}

Fig.~\ref{fig3} shows the average semantic SE versus $P_{\max}$ under different phase alignment accuracies for $\delta_\text{S}$ and $\delta_\text{B}$ in the single-waveguide PASS with $N\!=\!3$ and $D\!\!=\!\!20$\!\! m. Here, $\delta_\text{S}$ and $\delta_\text{B}$ are design parameters that quantify the allowable phase mismatch for the semantic and bit users, respectively. In practice, smaller values correspond to tighter phase alignment, whereas larger values relax the alignment requirement. The values 0.02, 0.5, and 100 used here can be interpreted as fine, moderate, and coarse phase alignment, respectively. It is notable that fine-tuning antenna positions with smaller values for $\delta_\text{S}$ and $\delta_\text{B}$, is critical for SE improvement. Tightening $\delta_\text{S}$ from a coarse value to a fine one yields the highest gains across the power range. More specifically, constraining $\delta_\text{S}$ with a tighter value offers the best performance as it directly boosts the objective function involving semantic SE. With $\delta_\text{S}=0.02$ and $\delta_\text{B}\!\!=\!\!100$, the curve trails the best-performing curve at low $P_{\max}$ values, but its performance improves significantly around 10 to 15\! dBm interval, and then closely matches the top curve at higher $P_{\max}$ values. By constrast, with $\delta_\text{S}\!=\!100$ and $\delta_\text{B}=0.02$, inferior performance is mainly due to the coarsely aligned $\delta_\text{S}$. When both users are coarsely aligned, i.e., $\delta_\text{S}=100$ and $\delta_\text{B}=100$, the semantic SE curve remains at the lowest level across all $P_{\max}$ values, negating the geometric advantages of PASS.

\begin{figure}[!t]
\centering
\includegraphics[trim={0cm 0cm 0cm 0cm},clip,width=1\columnwidth]{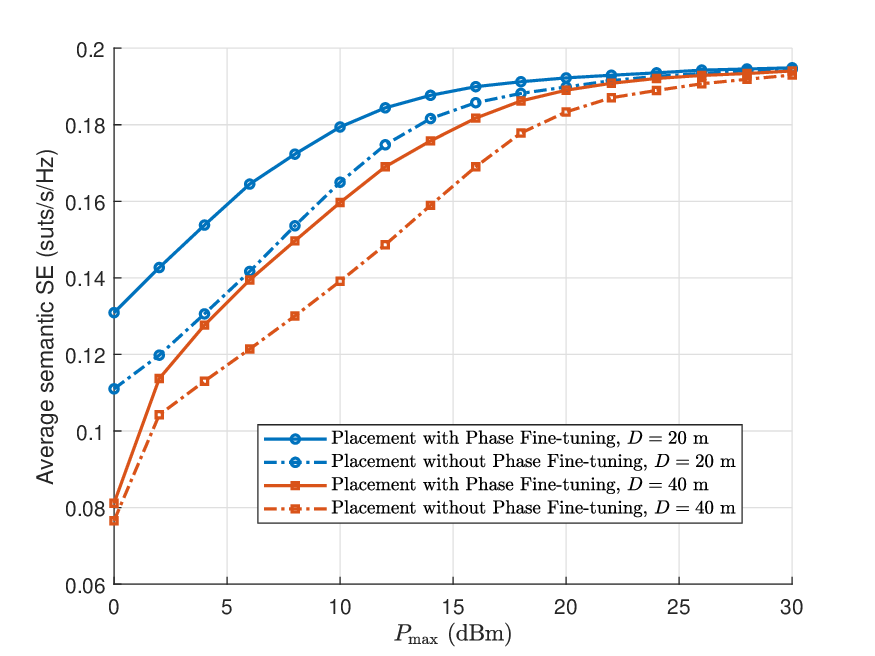}
\caption{Average semantic SE versus $P_{\max }$ for pinching antennas placement with and without phase fine-tuning in the single-waveguide PASS with $N\!=\!3$.}
\vspace{-9mm}
\label{fig4}
\end{figure}

Fig.~\ref{fig4} compares the schemes of pinching antennas placement with and without phase fine-tuning. In the former, the $N$ pinching antennas are further adjusted along the waveguide to improve phase alignment and enable more coherent combining at the semantic user. By contrast, in the scheme without phase fine-tuning, the antennas are simply placed with uniform $\lambda/2$ spacing starting from the antenna position that gives the highest channel gain for the semantic user. It is evident that phase fine-tuning offers higher semantic SE across entire $P_{\max}$ and for both simulated side lengths. For example, at $P_{\max}=10$\! dBm and $D\!=\!40$\! m, the phase fine-tuning placement offers an improvement of approximately 15\% in semantic SE over the scheme not invoking it.

\begin{figure}[!t]
\centering
\includegraphics[trim={0cm 0cm 0cm 0cm},clip,width=1\columnwidth]{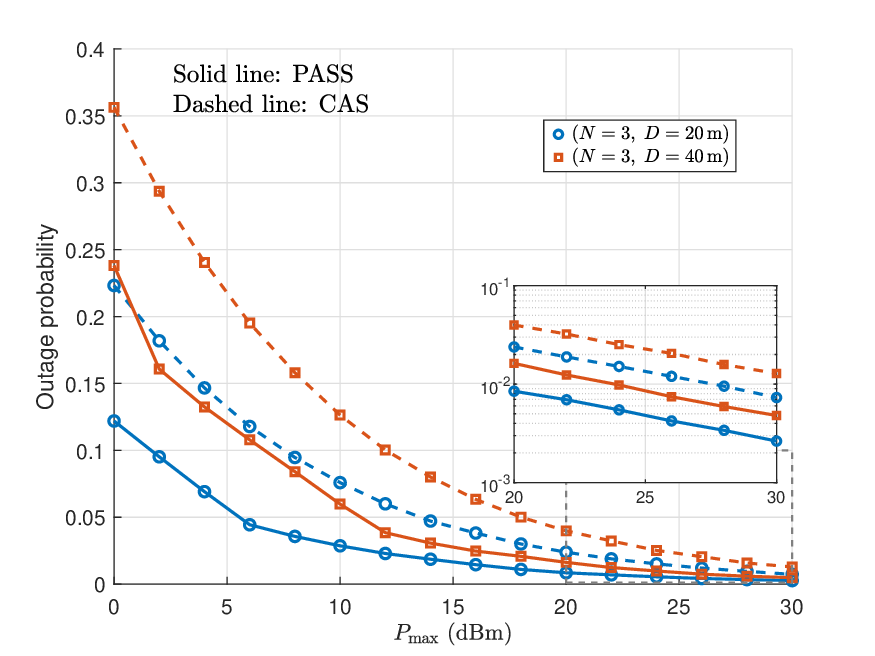}
\caption{Outage probability of the QoS and SIC feasibility versus $P_{\max}$ for the single-waveguide PASS and CAS.}
\vspace{-6mm}
\label{fig5}
\end{figure}

Fig.~\ref{fig5} compares the probability of the event when the bit-user QoS cannot be satisfied at both the users for the single-waveguide PASS and CAS. Across the entire power range, PASS yield a consistently lower outage than CAS. With flexible repositioning of antennas, PASS strengthen the effective channel gain towards the semantic user while maintaining the bit-user QoS and SIC-decodability. In contrast, due to the fixed antenna positions in CAS, random geometries are more likely to violate feasibility conditions. As $P_{\max}$ increases, all the curves decay monotonically, however PASS attain negligible outage at lower power levels, underscoring the reliability advantage in a heterogeneous users network.

\begin{figure}[!t]
\centering
\includegraphics[trim={0cm 0cm 0cm 0cm},clip,width=1\columnwidth]{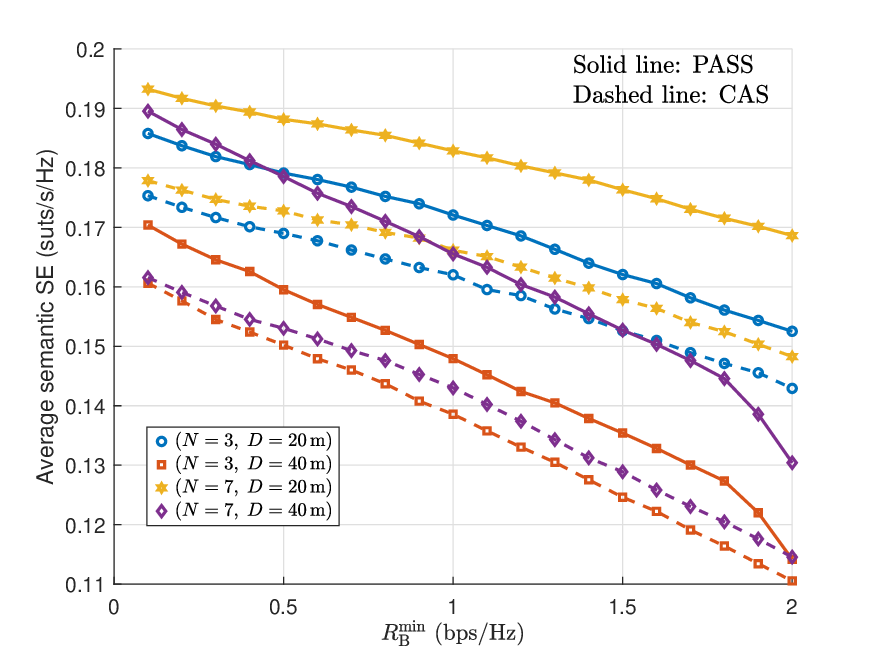}
\caption{Average semantic SE versus $R_{\text{B} }^{\text{min}}$ for the single-waveguide PASS and CAS at $P_{\max}\!=\!10$ dBm.}
\vspace{-9mm}
\label{fig6}
\end{figure}

Fig.~\ref{fig6} illustrates the average semantic SE performance for the compared single-waveguide PASS and CAS with varying bit-user QoS requirement. Both schemes exhibit a gradual reduction in semantic SE with increasing $R_{\text{B} }^{\text{min}}$ due to the higher bit-user resource allocation to satisfy its throughput requirement. However, PASS consistently achieve a higher semantic SE than CAS across the entire range for the simulated settings due to the flexibility of pinching antennas in creating favourable channel gains. Numerically, the average semantic SE gain for the single-waveguide PASS over CAS ranges from about 6\% to 10\% at $D=20$\! m, while at $40$\! m, it is between 7\% to 15\%, demonstrating the advantage of PASS across all $R_{\text{B} }^{\text{min}}$ values.

\begin{figure}[!t]
\centering
\includegraphics[trim={0cm 0cm 0cm 0cm},clip,width=1\columnwidth]{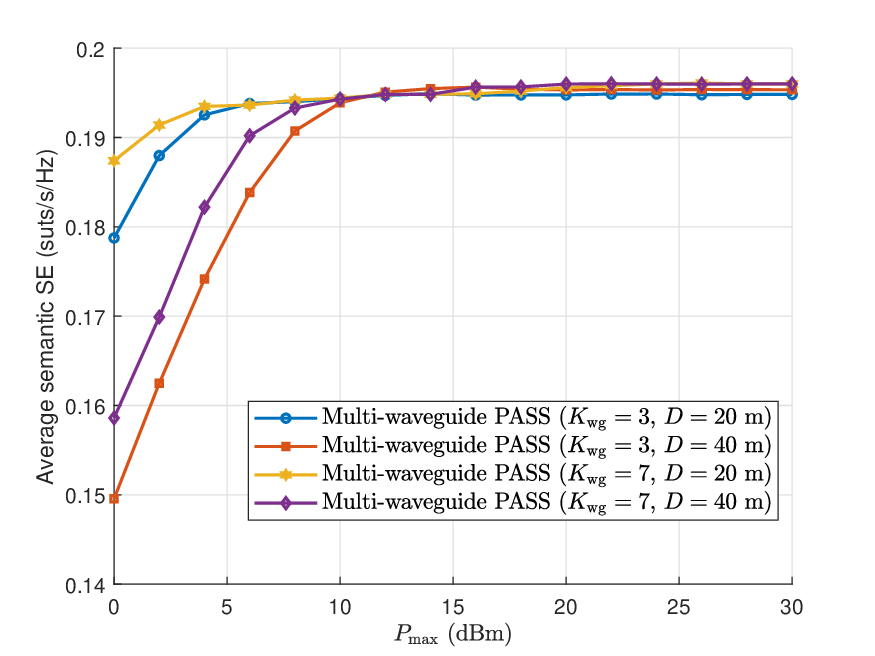}
\caption{Average semantic SE versus $P_{\max }$ for the NOMA assisted multi-waveguide PASS.}
\vspace{-9mm}
\label{fig7}
\end{figure}

Fig.~\ref{fig7} shows average semantic SE versus $P_{\max}$ for the multi-waveguide PASS. Consistent with the observations in Fig.~\ref{fig2}, a compact deployment area outperforms the larger deployment area due to reduced path loss and a stronger coherent combining effect, particularly for the low-to-moderate $P_{\max}$ values. Moreover, the multi-waveguide PASS offer improvement over a single-waveguide with multiple pinching antennas. This improvement stems from the additional spatial diversity offered by the multiple waveguides with fixed lateral offsets, relaxing the minimum antenna spacing constraint and enhancing the channel gain via flexible coherent combining. Particularly, the multi-waveguide PASS exhibit the largest performance advantage over the single-waveguide setup at $D\!=\!40$\! m and $P_{\max}\!\leq\!10$\!\! dBm for the same number of pinching antennas. As $P_{\max}$ increases, the performance gap in all settings gradually diminishes because the semantic similarity function approaches its upper bound.

\begin{figure}[!t]
\centering
\includegraphics[trim={0cm 0cm 0cm 0cm},clip,width=1\columnwidth]{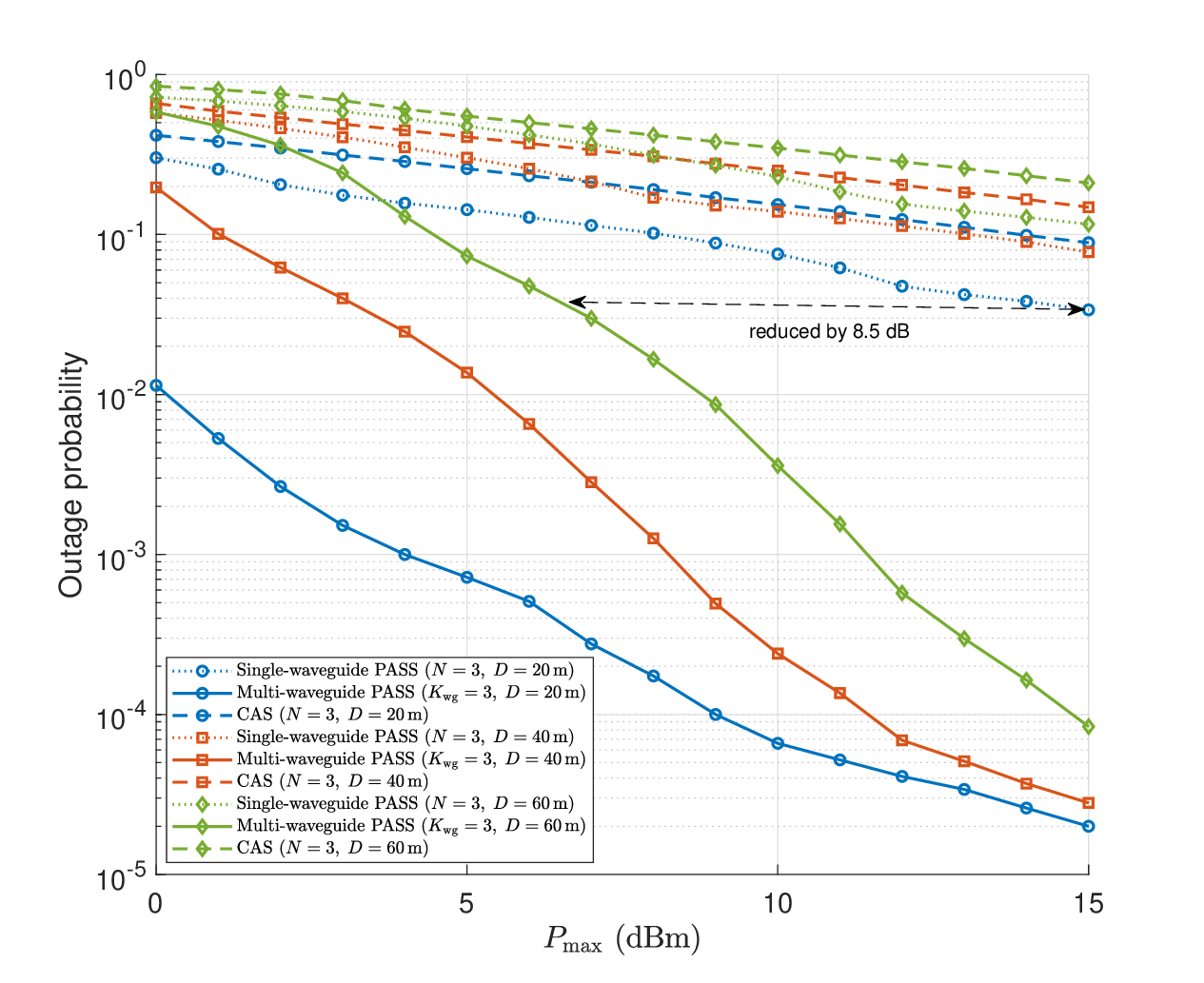}
\caption{Outage probability of the bit-user QoS and SIC feasibility versus $P_{\max }$ for the single-waveguide PASS, multi-waveguide PASS and CAS at $R_{\text{B} }^{\text{min}}\!=\!1.5$ bps/Hz.}
\vspace{-6mm}
\label{fig8}
\end{figure}

Fig.~\ref{fig8} compares the outage probability performance of the single-waveguide PASS, multi-waveguide PASS and CAS for the event when the bit-user QoS cannot be satisfied at both the users. For comparison, all architectures are evaluated with the same number of antennas. In contrast to Fig.~\ref{fig5}, we now consider more stringent conditions with wider user distribution areas and higher $R_{\text{B} }^{\text{min}}$. It is evident that the multi-waveguide PASS achieve superior performance throughout. Notably, the transmit power reduces by 8.5 dB under the stringent multi-waveguide setup compared to the single-waveguide PASS with three times smaller coverage area. This improvement is due to better mitigation of the unfavourable channel conditions by the multi-waveguide PASS, demonstrating suitability for heterogeneous users communication scenarios with wider coverage area and stricter bit-user QoS requirements. Conversely, when the system requirements are relaxed, the single‑waveguide PASS remain a viable low-complexity alternative.

\begin{figure}[!t]
\centering
\includegraphics[trim={0cm 0cm 0cm 0cm},clip,width=1\columnwidth]{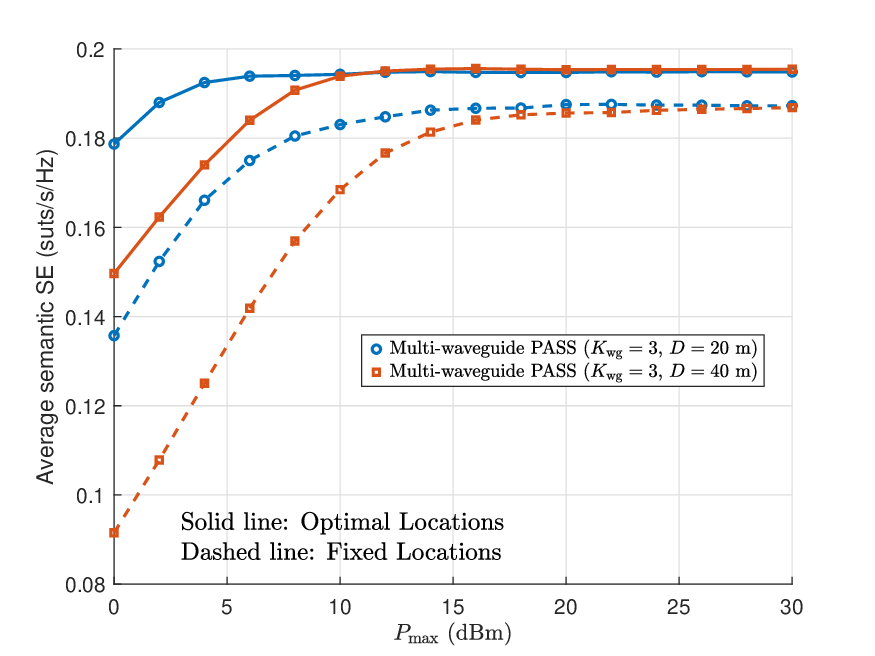}
\caption{Average semantic SE versus $P_{\max }$ for the fixed and optimal pinching locations in the multi-waveguide PASS.}
\vspace{-9mm}
\label{fig9}
\end{figure}

Fig.~\ref{fig9} compares the fixed and optimal pinching locations for the multi-waveguide scenario by plotting the average semantic SE versus $P_{\max}$ for both types. For the fixed locations configuration, we position a single pinching antenna at the centre of each waveguide in the multi-waveguide PASS. Notably, the improvement offered by the optimally located pinching antennas over the fixed positional pinching antennas is significantly higher for greater value of $D$ at low transmit power $P_{\max}\!\!=\!\!0$\! dBm, attesting the suitability of multi-waveguide PASS in low-power wider coverage scenarios. This improvement is attributed to the positional flexibility of pinching antennas via optimal placements in the multi-waveguide scenario, harnessing a strong effective channel gain for the heterogeneous users to maximize the semantic SE while meeting the bit-user QoS.

\begin{figure}[!t]
\centering
\includegraphics[trim={0cm 0cm 0cm 0cm},clip,width=1\columnwidth]{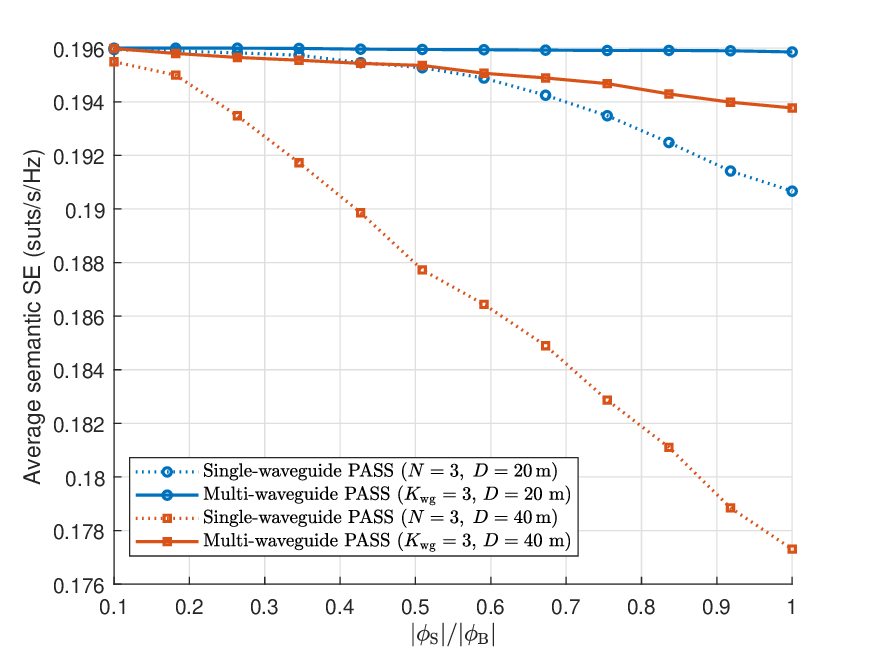}
\caption{Average semantic SE versus users distance ratio from the origin for the single-waveguide PASS and multi-waveguide PASS at $P_{\max}\!=\!10$ dBm.}
\vspace{-6mm}
\label{fig10}
\end{figure}

Fig.~\ref{fig10} shows the effect on PASS caused by varying the ratio of distance from the origin to the semantic and bit users. The averaged semantic SE here is obtained by grouping all realizations of the ratio $|\phi_\mathrm{S}|/|\phi_\mathrm{B}|$ within uniform ratio intervals and then performing averaging within each interval. As the ratio increases, the semantic and bit user starts to be at a comparable distance from the origin with the value 1 meaning that both are at equal euclidean distance. It is evident that the multi-waveguide PASS offer improvement over the singe-waveguide setup, specifically in the higher distance ratio regimes and at larger $D$. Compared to the single-waveguide PASS, there is an improvement of about 10\% for the multi-waveguide PASS at $D\!=\!40$\! m.

\begin{figure}[!t]
\centering
\includegraphics[trim={0cm 0cm 0cm 0cm},clip,width=1\columnwidth]{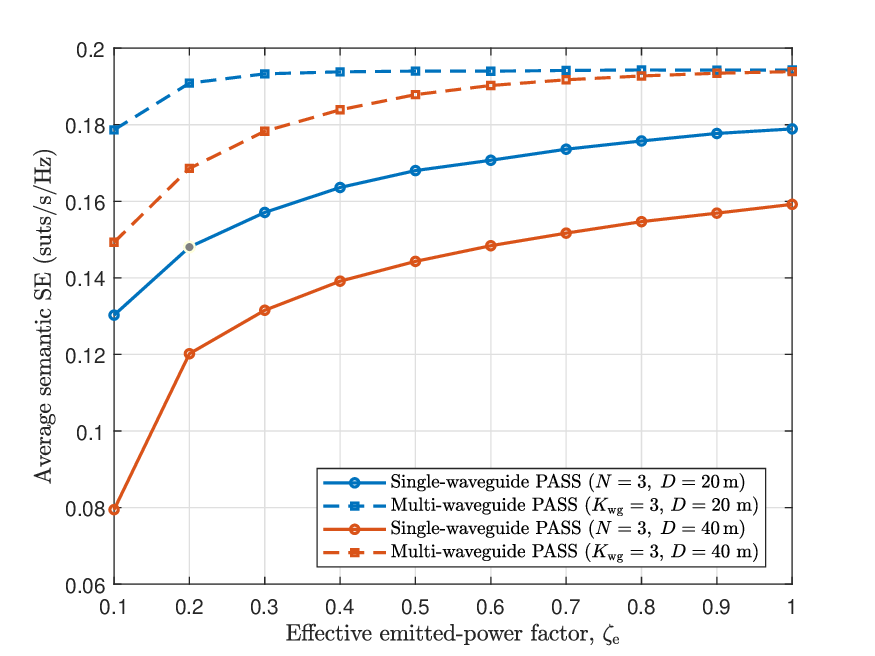}
\caption{Average semantic SE versus the effective emitted-power factor $\zeta_{\mathrm{e}}$.}
\vspace{-9mm}
\label{R1-1}
\end{figure}

\color{blue}
To examine the robustness of the proposed framework under non-ideal emitted-power conditions, Fig.~\ref{R1-1} plots the average semantic SE versus the effective emitted-power factor. The value $\zeta_{\mathrm{e}}=1$ corresponds to the ideal equal-emission model, while smaller values represent reduced effective emitted power caused by non-ideal coupling/radiation. For both single-waveguide and multi-waveguide PASS, the semantic SE decreases as $\zeta_{\mathrm{e}}$ decreases. However, the relative performance advantage of the multi-waveguide PASS model remains consistent, showing that the proposed design is robust to moderate emitted-power loss.

\begin{figure}[!t]
\centering
\includegraphics[trim={0cm 0cm 0cm 0cm},clip,width=1\columnwidth]{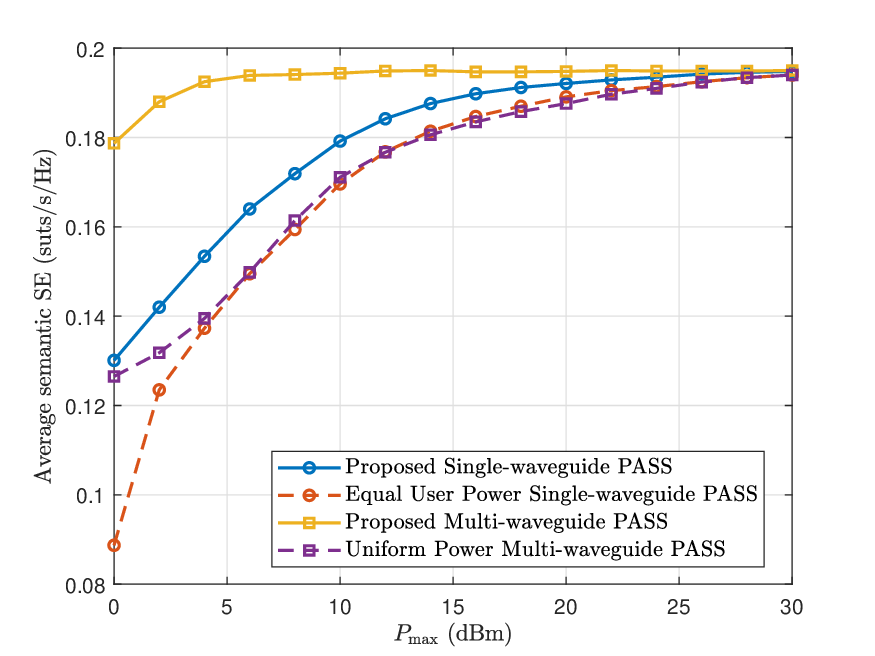}
\caption{Average semantic SE comparison of the proposed single-waveguide and multi-waveguide PASS with heuristic baselines for $N=K_{\mathrm{wg}}=3$, and $D=20$ m.}
\vspace{-6mm}
\label{R1-2}
\end{figure}

To validate the effectiveness and stability of the proposed approach, we compare it against the heuristic baselines as shown in Fig.~\ref{R1-2}. For the single-waveguide PASS model, we compare the proposed closed-form NOMA power-allocation design with a fixed equal user power allocation. Both the single-waveguide schemes use the same antenna-positioning procedure and the same QoS feasibility checks. For the multi-waveguide case, the heuristic approach uses uniform waveguide power splitting instead of the proposed MM-based waveguide power optimization. It is evident that the proposed single-waveguide and multi-waveguide algorithms achieve higher average semantic SE than the corresponding baselines. This confirms that the performance improvement comes from the proposed feasibility-preserving optimization framework, rather than simply from using PASS.

\begin{figure}[!t]
\centering
\includegraphics[trim={0cm 0cm 0cm 0cm},clip,width=1\columnwidth]
{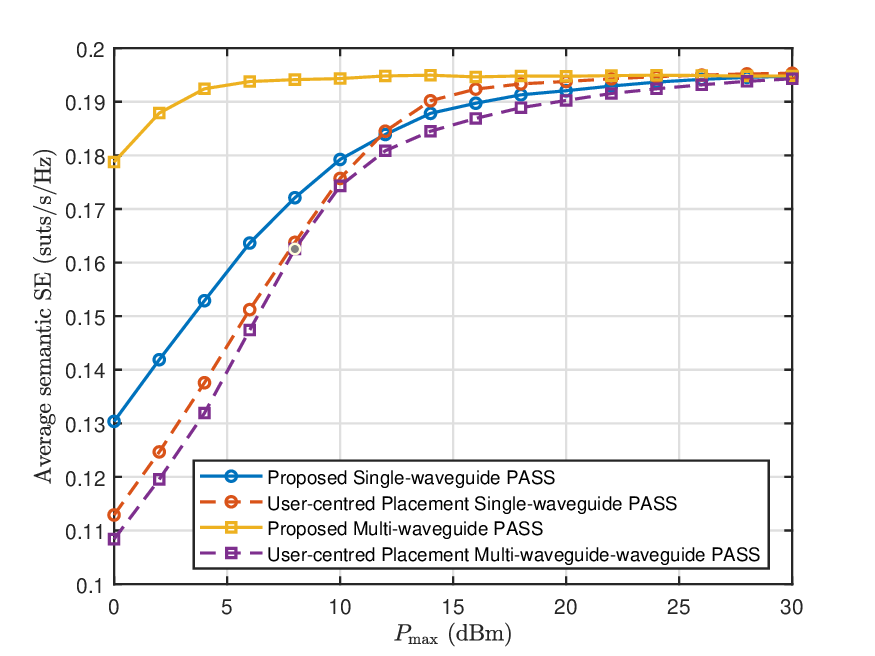}
\caption{Average semantic SE comparison of the proposed PASS optimization algorithms with semantic-user-centric antenna-placement heuristic baselines for $N=K_{\mathrm{wg}}=3$, and $D=20$ m.}
\vspace{-9.5mm}
\label{R1-3}
\end{figure}

To further validate the benefit of the proposed optimization framework, we consider a semantic-user-centric pinching antenna placement heuristic baseline in Fig.~\ref{R1-3}. The heuristic approach here exploits the basic PASS intuition that placing the radiation point closer to the intended user reduces the free-space propagation distance. Unlike the proposed method, the heuristic does not perform feasibility-preserving bisection procedure, phase refinement, or waveguide power-splitting optimization. This demonstrates the effectiveness of the proposed framework against the heuristic baselines.

\color{black}
\vspace{-0.6599em}
\section{Conclusions}
This paper examined PASS for the single-waveguide and multi-waveguide scenarios in a heterogeneous users NOMA framework to maximize the semantic SE subject to the bit-user QoS. For the single-waveguide PASS, the joint optimization problem of users power allocation coefficients and pinching antennas position was decoupled into two subproblems and solved using an AO method. For fixed positions of pinching antennas, the optimized power allocation coefficient for the semantic user was obtained by solving the users power allocation subproblem under the prescribed bit-to-semantic decoding and QoS requirements. The optimal positions of pinching antennas were then determined using an iterative one-dimensional bisection strategy under the minimum spacing constraint. Performance improved with the number of antennas, and this gain became more pronounced in the single-waveguide setup under phase alignment. For the multi-waveguide scenario, the waveguide power allocation subproblem was solved using the MM strategy. Simulation results demonstrate that the multi-waveguide PASS outperform both the single-waveguide PASS and fixed location baselines, while also exhibiting a lower outage probability under stringent conditions. These findings highlight the potential of PASS as a promising technology for enabling heterogeneous users wireless networks.


\color{blue}
\vspace{-0.8em}
\appendix[Extension to Clustered Multi-User NOMA]
For a practical clustered multi-user extension, a higher-layer scheduler can form different cluster types according to user demands, channel conditions, QoS requirements, and available time-frequency resource. Bit-users only clusters will follow conventional NOMA decoding, semantic-users only clusters will utilize channel reconstruction-based SIC \cite{chen2023uplink}, and heterogeneous semantic-bit clusters will benefit from the bit-to-semantic decoding strategy \cite{mu2022heterogeneous}. The transmitted signal for a heterogeneous cluster $g$ is
\begin{equation}
s_g=\sqrt{\alpha_{\text{B}, g} P_g} s_{\text{B}, g}+\sqrt{\alpha_{\text{S}, g} P_g} s_{\text{S}, g},
\end{equation}
where $P_g$ denotes the transmit power allocated to cluster $g$, $\alpha_{\text{B}, g}$ and $\alpha_{\text{S}, g}$ are the power-allocation coefficients for the bit and semantic users, respectively satisfying  $\alpha_{\text{B}, g}+\alpha_{\text{S}, g}=1$, and $s_{\text{B}, g}$ and $s_{\text{S}, g}$ denote the signals intended for these users.  Let $\mathcal{G}=\{1, \ldots, G\}$ is the set of heterogeneous clusters and assuming that there is no inter-cluster interference, then the corresponding optimization problem can be expressed as
\begin{equation}
\max_{\mathbf{x}_g,\, \alpha_{\text{S},g},\, \boldsymbol{\beta}_g}
\sum_{g \in \mathcal{G}} R_{\text{S},g},
\end{equation}
with per-cluster QoS, SIC, NOMA decoding order, power-budget, and PASS deployment constraints. This shows that the proposed two-user model can serve as a building block for larger heterogeneous semantic-bit NOMA systems.

However, a multi-user NOMA formulation in which multiple semantic and bit users share the same resource would lead to a mixed-integer non-convex problem due to users grouping and decoding-order variables. This would introduce longer SIC chains and increase the risk of SIC error propagation. Such a design is beyond the scope of the present work.
\color{black}
\bibliographystyle{IEEEtran}
\color{black}
\bibliography{references}
\end{document}